\documentclass[5p,final]{elsarticle}
\usepackage{booktabs}
\usepackage{threeparttable}

\usepackage{lineno,hyperref}
\modulolinenumbers[5]
\usepackage{bm}
\usepackage{amsmath}

\journal{Nuclear Instruments and Methods in Physics Research A}

\begin{document}

\begin{frontmatter}

\title{Apparatus for in-beam hyperfine interactions and $g$-factor measurements: design and operation}


\author[ANUresearch]{A.E.~Stuchbery \corref{mycorrespondingauthor}}
\cortext[mycorrespondingauthor]{Corresponding author}
\ead{andrew.stuchbery@anu.edu.au}
\author[ANUresearch]{A.B.~Harding \fnref{myfootnote}}
\author[ANUresearch]{D.C.~Weisser \fnref{myfootnote}}
\fntext[myfootnote]{Deceased.}
\author[ANUresearch]{N.R.~Lobanov}

\address[ANUresearch]{Department of Nuclear Physics,
Research School of Physics, The~Australian National
University, Canberra, ACT~2601, Australia}





\begin{abstract}
The design and operation of apparatus for measurements of in-beam hyperfine interactions and nuclear excited-state $g$~factors is described. This apparatus enables a magnetic field of about 0.1 tesla to be applied to the target and the target temperature to be set between $\sim 4$~K and room temperature. Design concepts are developed mainly in terms of transient-field $g$-factor measurements following Coulomb excitation by the implantation perturbed angular correlation (IMPAC) technique. The formalism for perturbed angular correlations is outlined and a figure of merit for optimizing these measurements is derived to inform design.  Particle detection is based on the use of silicon photodiodes of rectangular shape. The particle-$\gamma$ angular correlation formalism for this case is described.  The experimental program to date includes temperature-dependent studies of hyperfine fields, transient-field $g$-factor measurements, and time-dependent perturbed angular distribution (TDPAD) studies.
\end{abstract}



\end{frontmatter}


\section{Introduction} \label{Intro}

It is a practical imperative that hyperfine magnetic fields of the
order of 10 to 1000 tesla must be applied to the nucleus in order to
measure the $g$~factor (or magnetic moment) of a short-lived excited state with a
lifetime less than a few nanoseconds. Such fields are experienced at
the nuclei of ions implanted into ferromagnetic hosts. Ions in
motion through a polarized ferromagnetic medium experience a
\textit{transient hyperfine field} \cite{ben80}, whereas ions at rest
experience a \textit{static hyperfine field} \cite{Krane1983}.

The implantation perturbed angular correlation (IMPAC) technique has
proved very useful for measurements of magnetic moments \cite{ben80,speidel02,benczerkoller07}
and, where nuclear moments are known, as a technique to study
condensed matter problems \cite{Kugel-PhysRevB.13.3697,recknagel75} and ion-solid interactions
\cite{Speidel1997,Stuchbery.PhysRevLett.82.3637,SPEIDEL2004604}.

This paper describes apparatus designed in the Department of Nuclear Physics at the Australian National University (ANU), and operated in the Heavy Ion Accelerator Facility (HIAF) at the Australian National University for about a decade. Named Hyperion (hyperfine interactions on-line), or referred to as the Hyperfine Spectrometer, the apparatus was designed to study in-beam hyperfine interactions by perturbed angular correlation (PAC) techniques, with a particular emphasis on the IMPAC technique applied to excited nuclear states with lifetimes of the order of picoseconds. It incorporates a number of novel design features and gives control of the target temperature between $\sim 4$~K and room temperature, which enables temperature-dependent studies of hyperfine interactions in a range of magnetic materials.

The paper is set out as follows. Section \ref{sect:technique} outlines the IMPAC technique. Section \ref{sect:expt-design-opt} focuses on the optimization of experiments to inform the design of the apparatus (i.e. the Hyperfine Spectrometer). The Hyperfine Spectrometer is described and its design features discussed in section \ref{sect:design}. Operational aspects and performance are discussed in section~\ref{sect:performance}. Section ~\ref{sect:commissioning} describes the commissioning experiments, which exploited the temperature control features of the apparatus. A summary of the $g$-factor measurements that have been performed with the Hyperfine Spectrometer using the transient-field technique is presented in section~\ref{sect:TF}, along with a discussion of some novel features of these experiments. Examples of time-dependent perturbed angular distribution (TDPAD) measurements that have been performed are given in section~\ref{sect:TDPAD}. A summary and outlook follow. At several points in the text particle detection based on rectangular photodiodes is discussed. An appendix gives some technical details on the computation of the particle-$\gamma$ angular correlations following Coulomb excitation for these rectangular particle detectors.

\section{Technique Outline}
\label{sect:technique}

\subsection{Perturbed angular correlations: the IMPAC technique}
\label{sect:IMPAC}

Perturbed angular correlation techniques cover a wide range of methodologies that can include time-integral or time-differential measurements, and angular correlations between particles and $\gamma$~rays or $\gamma\gamma$ angular correlations \cite{recknagel}. The ANU hyperfine spectrometer was designed with many of these variations in mind, but with an emphasis on integral perturbed particle-$\gamma$ angular correlations following Coulomb excitation - the typical IMPAC measurement. Options were kept open for applications including $\gamma\gamma$ angular correlations and time-differential measurements, but generally speaking, if the apparatus is designed for IMPAC measurements following Coulomb excitation, it will also be useful for the other types of PAC measurements as well.

\begin{figure}
\begin{center}
\resizebox{8.5cm}{!}{\includegraphics{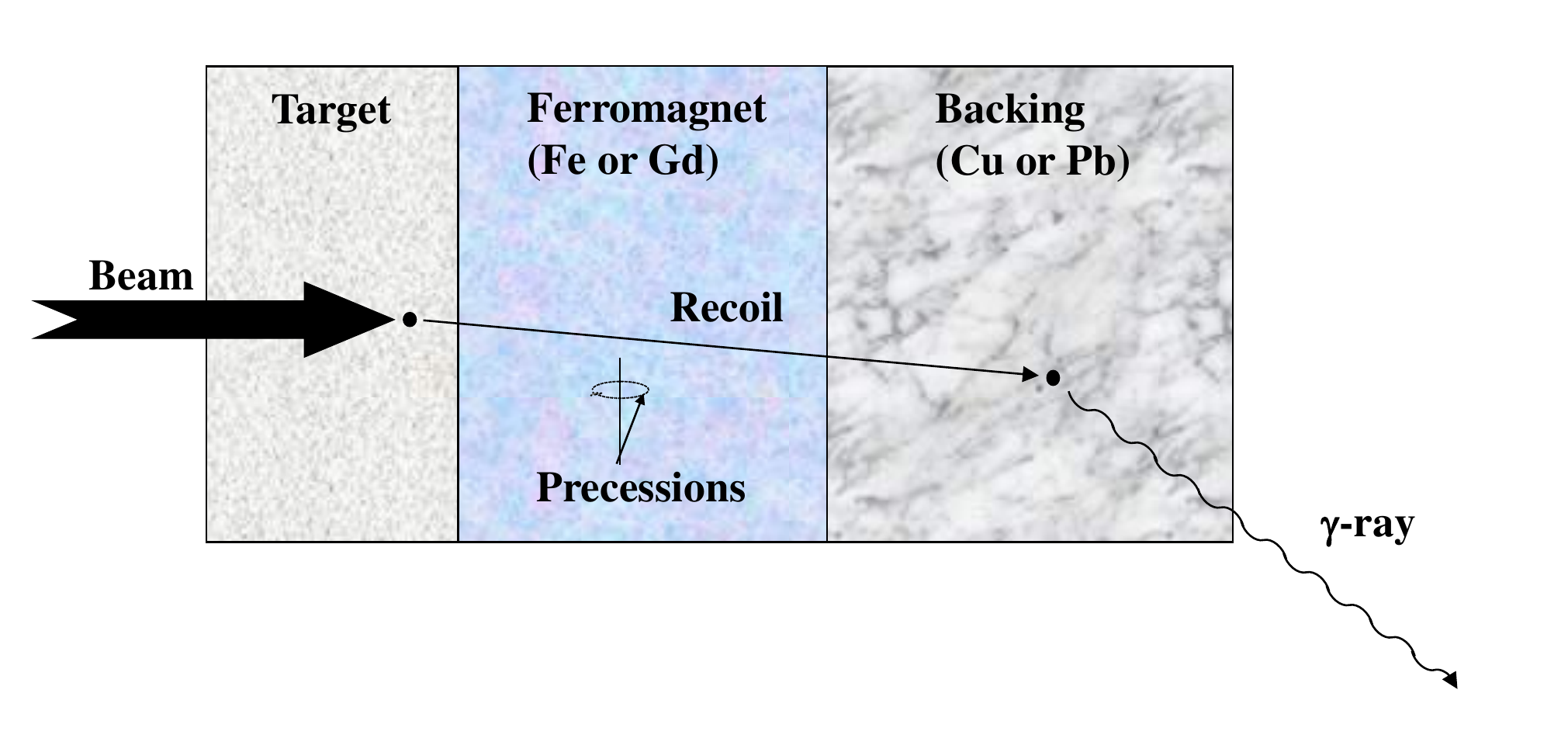}}
\caption{Representation of the triple-layered target used in conventional `thin-foil' transient-field $g$-factor measurements \cite{ben80}. The target nuclei of interest are Coulomb excited by the beam and recoil through the ferromagnetic layer where they experience the transient field, before coming to rest in a non-magnetic backing layer (often copper or lead). (Diagram from Ref.~\cite{MPRthesis}.)} \label{fig:target}
\end{center}
\end{figure}

The `classic' IMPAC measurements, performed since the 1960s, used `conventional kinematics', i.e. a lower mass beam was employed to Coulomb excite a heavier target nucleus. De-excitation $\gamma$ rays were detected in coincidence with backscattered beam ions detected in an annular counter placed around the beam axis \cite{ben80}. More recently there has been a shift to `inverse' kinematics in which a heavier beam ion is Coulomb-excited on a lower-mass target with the knock-on target ions detected at forward angles \cite{benczerkoller07}. These `conventional' and `inverse' kinematics experiments both correspond to near head-on collisions in the centre-of-mass frame. The advantages of inverse kinematics over conventional kinematics for IMPAC measurements include (i) the improved sensitivity achieved by virtue of the forward-focusing of the reaction products in the laboratory frame (the particle detector therefore can cover a larger solid angle in the centre-of-mass frame), and (ii) the applicability of inverse kinematics to experiments on exotic nuclei produced as radioactive beams, which open up new regimes for nuclear structure studies \cite{Kr76_KUMBARTZKI2004p213,Te132TF_BENCZERKOLLER2008p241,Sn126TF_PhysRevC.86.034319,Zn72_LVTF_PhysRevC.89.054316}. A third alternative for the reaction is to excite the beam ions in glancing collisions on heavier target nuclei \cite{Speidel1991,CUB1992p304,STUCHBERY2005p81,S38-40_PhysRevLett.96.112503,S38-40_PhysRevC.74.054307,Zn72_PhysRevC.85.034334}. Further discussion of conventional versus inverse kinematics measurements may be found in the reviews of Speidel {\em et al.} \cite{speidel02} and Benczer-Koller and Kumbartzki \cite{benczerkoller07}.

In parallel with the shift from conventional to inverse kinematics, there has also been a move to the use of inexpensive silicon photodiodes to replace silicon surface barrier detectors in many applications. The apparatus described here was designed for use with silicon photodiodes, which have a rectangular shape, rather than the traditional surface barrier detectors having cylindrical symmetry. For the applications of interest, breaking the cylindrical symmetry of the particle detection means that the azimuthal symmetry about the beam axis is broken. The formalism for the perturbed angular correlation becomes a little more complex, as discussed below. (See section \ref{sect:ACcalculation} and \ref{sect:rectangular_tensors}). 

\subsection{Transient-field IMPAC}
\label{sect:TF_IMPAC}

Figure \ref{fig:target} represents a three-layer target as employed in `thin-foil' transient-field $g$-factor measurements. (See Ref.~\cite{ben80} for a description of thin-foil versus thick-foil measurements. Today essentially all transient-field measurements are thin-foil measurements using a triple-layer target like that represented in Fig.~\ref{fig:target}.) The nuclei of interest in the first layer of the target are Coulomb excited and recoil into a suitable polarized ferromagnetic host (the second layer of the target). As they move through this ferromagnetic layer, the transient field acts on the nuclei of the ions, causing the nuclear spin to precess about the direction of the magnetic field. The ions slow and then stop in the third, nonmagnetic layer of the target, where the nuclear decay (predominantly) takes place. The precession of the nuclear spin is observed via the perturbed particle-$\gamma$ angular correlation of the de-excitation $\gamma$ rays.

\begin{figure}
\begin{center}
\resizebox{8.5cm}{!}{\includegraphics{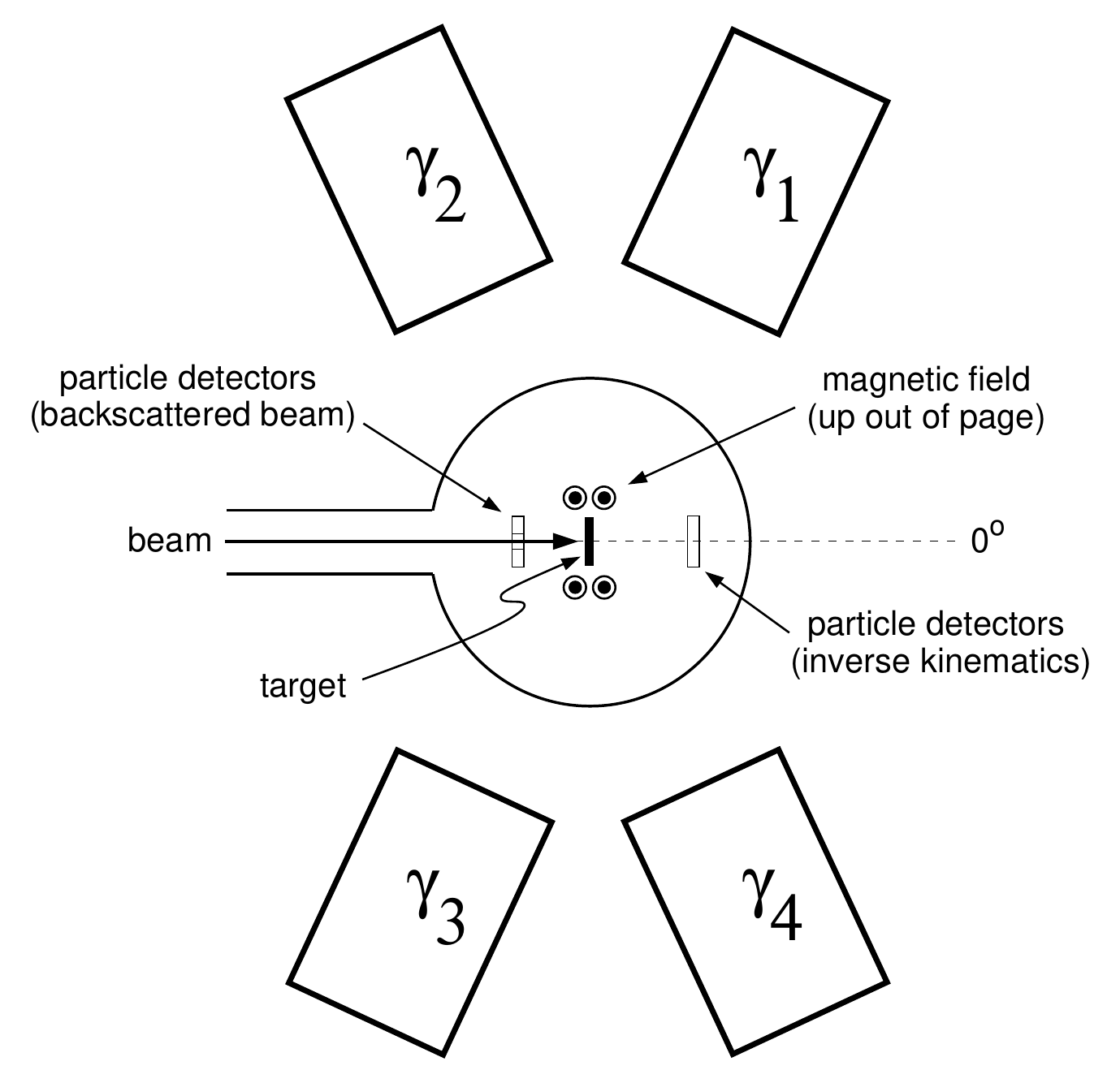}} \caption{Plan-view sketch of the apparatus used for transient-field $g$-factor measurements. The $\gamma$-ray detectors are in the horizontal plane through the beam axis and the magnetic field applied to the target is in the vertical plane.} \label{fig:chamber-sketch}
\end{center}
\end{figure}

The arrangement of the vacuum chamber, the magnetic field applied to the target, and the particle and $\gamma$-ray detectors, is shown schematically in Fig.~\ref{fig:chamber-sketch}. This sketch represents a view from above. Gamma-ray detectors are in the horizontal plane, magnetic field `up' is vertically upwards, and the beam is along the $z$-axis ($\theta=0^{\circ}$).

In a typical transient-field measurement the precession angle of the nucleus is small. The effect of the transient field is simply to
rotate the angular correlation pattern $W(\theta)$ through an angle $\Delta \theta$ about the applied field direction. For
a {\em thin-foil} transient-field measurement, using a target as represented in Fig.~\ref{fig:target}, the
precession angle is
\begin{equation}
\Delta \theta = - g \frac{\mu_{\rm N}}{\hbar} \int^{t_e}_{t_i}
B_{\rm tf}(t) e^{-t/\tau} dt,
\end{equation}
where $g$ is the nuclear $g$~factor, $\mu_{\rm N}$ is the nuclear magneton, $\hbar$ is the reduced Planck constant, and $B_{\rm tf}$ is the transient-field strength, which depends on the time $t$ through its dependence on the velocity of the ion as it slows in the ferromagnetic layer of the target. The factor $e^{-t/\tau}$ accounts for nuclear decays during transit through the ferromagnetic foil, but typically remains near unity because the excited nucleus usually traverses a thin ferromagnetic foil in a time ($ < 1$~ps) that is short compared to the mean life of the excited state of interest. The time at which the ion enters into the ferromagnetic layer is $t_i$ and the time at which it exits is $t_e$.

For the magnetic field in the `up' (`down') direction the perturbed angular correlation is
\begin{equation}
W^{\uparrow(\downarrow)}(\theta) = W(\theta \mp \Delta \theta) \simeq
W(\theta) \mp \Delta \theta \frac{dW}{d \theta},
\end{equation}
where the negative sign applies for field up ($\uparrow$) and the
positive sign applies for field down ($\downarrow$). The meaning of
`up' and `down' is defined in Fig.~\ref{fig:chamber-sketch} and in Fig.~\ref{fig:particle-angles}. The precession angle is determined
experimentally by placing a pair of $\gamma$-ray detectors at angles
$\pm \theta_\gamma$ with respect to the beam axis, and forming a
double ratio
\begin{equation} \label{eq:rho}
\rho =
\sqrt{\frac{N^{\uparrow}(+\theta_\gamma)}{N^{\downarrow}(+\theta_\gamma)}
\frac{N^{\downarrow}(-\theta_\gamma)}{N^{\uparrow}(-\theta_\gamma)}},
\end{equation}
where, for example, $N^{\uparrow}(+\theta_\gamma)$ is the number of
counts recorded in the detector at $+\theta_\gamma$ for field up. Note that in spherical polar coordinates as defined in Fig.~\ref{fig:particle-angles}, the detector designated at angle $-\theta_\gamma$ is at $(\theta, \phi)= (\theta_\gamma, 180^{\circ}$).

Defining
\begin{equation} \label{eq:epsilon}
\epsilon = \frac{1 - \rho}{1 + \rho},
\end{equation}
the experimental precession angle is given by
\begin{equation}\label{eq:delta-theta}
\Delta \theta = \epsilon/S,
\end{equation}
where
\begin{equation}\label{eq:S}
S = S(\theta_{\gamma}) = \left . \frac{1}{W}\frac{dW}{d\theta} \right |_{\theta_\gamma}.
\end{equation}

The `effect' $\epsilon$ is formally equivalent to
\begin{eqnarray}
\epsilon &=& \frac{N^{\downarrow}(+\theta_\gamma)-N^{\uparrow}(+\theta_\gamma)}{N^{\downarrow}(+\theta_\gamma)+N^{\uparrow}(+\theta_\gamma)}  \\
         &=& \frac{W^{\downarrow}(+\theta_\gamma)-W^{\uparrow}(+\theta_\gamma)}{W^{\downarrow}(+\theta_\gamma)+W^{\uparrow}(+\theta_\gamma)}.
\end{eqnarray}
%

A typical precession measurement usually requires a relatively long run with a pair (or pairs) of $\gamma$-ray detectors at angles $\pm \theta_\gamma$ (and $180^\circ \pm \theta_\gamma$) to determine the `effect' $\epsilon$, along with a somewhat shorter run to determine the `slope' $S(\theta_{\gamma})$. For `safe' Coulomb excitation measurements \cite{Cline}, and with the excited nuclei implanted into a medium that does not perturb the spin alignment (typically a cubic metal), the angular correlation, and hence $S(\theta_{\gamma})$, can be calculated precisely; for examples see Refs.
\cite{BOLOTIN1983,Stuchbery1985,STUCHBERY1985-Os,BYRNE1987,STUCHBERY1988,LAMPARD1989,Doran.PhysRevC.40.2035,Stuchbery1991,STUCHBERY1991a,Stuchbery1992,Stuchbery1994,LAMPARD1994,ANDERSSEN1995,STUCHBERY1998,ROBINSON1999,STUCHBERY2000,BEZAKOVA2000,Mantica.PhysRevC.63.034312,East.PhysRevC.79.024303,Chamoli.PhysRevC.80.054301,Chamoli.PhysRevC.83.054318}. One caveat on this statement is that $S(\theta_{\gamma})$ is a very sensitive function of $\theta_{\gamma}$ when the detectors are placed near the maximum slope of the angular correlation as is done for precession measurements. In such cases, if there is an offset in the $\gamma$-ray detector angle from its nominal value, the calculated slope will be inaccurate. Such inaccuracies largely factor out if the experiment is designed to measure the relative precessions of two or more states simultaneously under the same experimental conditions; however the detector angle must be known accurately to calculate $S(\theta_{\gamma})$ in cases where the absolute magnitude of the precession angle is required.

\subsection{Angular correlation calculation}
\label{sect:ACcalculation}

\begin{figure}
\begin{center}
    \resizebox{8.5cm}{!}{\includegraphics{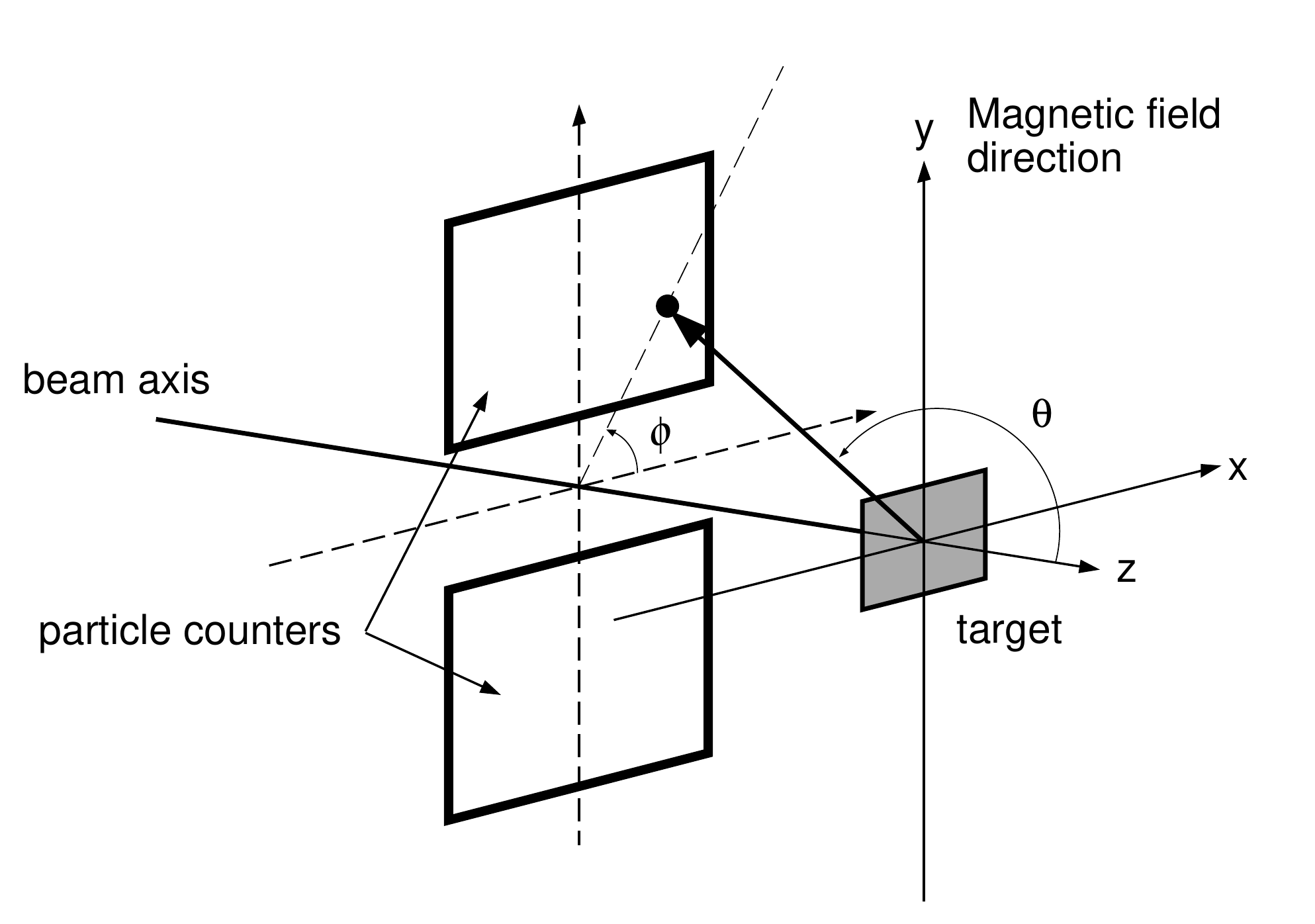}}
\caption{Particle detector schematic for backscattering reactions indicating the beam direction ($z$-axis)
and magnetic field direction ($y$-axis). The spherical polar angles $(\theta,\phi)$ are shown. Gamma-ray detectors are centred in the horizontal $xz$ or $\phi=0$ plane.  For convenience $\gamma$-ray detectors having $\phi=180^{\circ}$ are often designated as being at $-\theta_{\gamma}$.}
\label{fig:particle-angles}
\end{center}
\end{figure}

The particle-$\gamma$ angular correlation in general takes the form (see Refs.~\cite{PDCO-I.Stuchbery2002753,ADpaper.STUCHBERY200369,AW}
and references therein)
\begin{equation}
W(\theta_p, \theta_\gamma, \Delta \phi) = \sum_{k q} B_{k
q}(\theta_p) Q_k F_k D^{k *}_{q 0}(\Delta \phi, \theta_\gamma,
0), \label{eq:pac}
\end{equation}
where $(\theta_p, \phi_p)$ and $(\theta_\gamma, \phi_\gamma)$ are the spherical polar angles corresponding to particle and $\gamma$-ray detection, respectively, with the $z$-axis along the beam direction, and $\Delta \phi
= \phi_\gamma - \phi_p$. The statistical tensor $B_{kq}(\theta_p)$ defines the spin alignment of the initial state. $F_k$ represents the usual $F$-coefficient for the $\gamma$-ray transition, which depends on the initial and final level spins and the (possibly mixed) multipolarity of the transition. $Q_k$ is the attenuation factor for the finite size of the $\gamma$-ray detector and $D^{k *}_{q 0}(\Delta \phi, \theta_\gamma, 0)$ is the rotation matrix. In most of the applications of interest we are concerned with either $E2$ transitions or mixed $M1+E2$ transitions for which $k =0,2,4$.

\begin{figure*}[t]
\begin{center}
\resizebox{14cm}{!}{\includegraphics{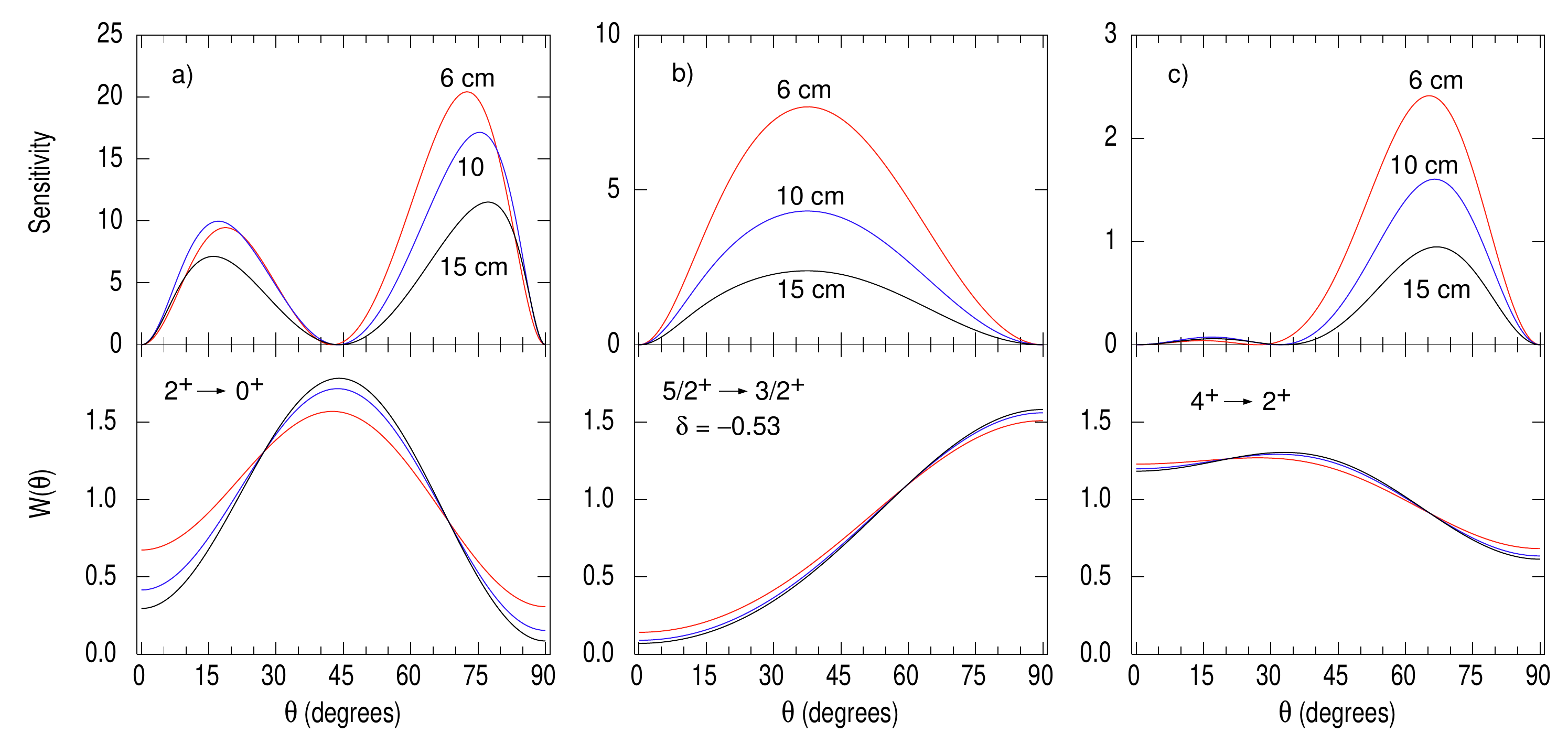}} \caption{Angular correlations and the corresponding sensitivity plots based on Eq.~(\ref{eq:S2N}) for (a) a typical $2^+ \rightarrow 0^+$ transition, (b) a mixed $5/2^+ \rightarrow 3/2^+$ transition (see Ref.~\cite{Chamoli.PhysRevC.80.054301}), and (c) a typical $4^+ \rightarrow 2^+$ transition, all following Coulomb excitation and detection of backscattered beam ions in a pair of rectangular particle detectors approximately 3.8 mm above and below the beam axis,
10 mm high, 9 mm wide and 16 mm back from the target. The Ge detector crystal is taken to have 7 cm diameter and 7 cm length. Three target to detector-face distances are shown (6, 10 and 15 cm). The vertical scale of the sensitivity plot is arbitrary but equivalent for all three transitions; note however that the scale changes.}
\label{fig:s2N}
\end{center}
\end{figure*}

For applications of the angular correlation formalism to experiments with the Hyperfine Spectrometer, the beam defines the $z$-axis, the $y$-axis, which is also the positive magnetic field direction, is vertically upwards, and the $x$-axis (or $\phi=0$ in spherical polar co-ordinates) is in the horizontal plane. Figure~\ref{fig:particle-angles} shows the relevant axes and angles for a pair of particle detectors and conventional backscattering kinematics. The finite size of the particle detectors is taken into account by integrating the statistical tensors over the acceptance of the particle detectors. Details of the procedure are given in \ref{sect:rectangular_tensors}. Once this integration has been performed, $W(\theta_p, \theta_\gamma, \Delta \phi)$ can be evaluated from Eq.~(\ref{eq:pac}), with $\Delta \phi = \phi_\gamma - \phi_p = \pi/2$.

\section{Optimization and design of experiment}
\label{sect:expt-design-opt}

\subsection{Figure of Merit}
\label{sect:IMPAC-FOM}

To design the apparatus, it is useful to define a figure of merit for IMPAC measurements that indicates the relative experimental precision achieved as a function of the $\gamma$-ray and particle-detector angles and solid angles. These considerations impact on the design of the apparatus in terms of the space allocated for particle detectors, the shape and position of the magnet return yoke, the design of the target mount, and generally where $\gamma$-ray absorption must be minimized.

If we consider a pair of $\gamma$-ray detectors at angles $\pm \theta_\gamma$ and make the assumptions that
(i) $ N^{\uparrow}(+\theta_\gamma) \approx N^{\downarrow}(+\theta_\gamma) \approx N^{\downarrow}(-\theta_\gamma)  \approx N^{\uparrow}(-\theta_\gamma) \approx N$, and
(ii) the peak to background in the $\gamma$-ray detector is high, then based on equations (\ref{eq:rho}) - (\ref{eq:S}), the statistical uncertainty on $\Delta \theta$ is
\begin{equation}
\sigma_{\Delta \theta} \approx \frac{1}{2} \frac{1}{S\sqrt{N}}.
\end{equation}
Thus to minimize the uncertainty on the measured precession angle requires that $S\sqrt{N}$ be maximized. A figure of merit can thus be defined as $S^2N$ so that
optimal experimental conditions occur for the maximum value of $S^2N$. This form for the figure of merit was chosen because it scales with the number of counts achieved, and thus with the number of detector pairs included in the experiment, and with the beam time allocated to the measurement. It is useful to note that $S^2N \propto S^2W \propto \frac{1}{W} (\frac{dW}{d\theta})^2$.

Although assumption (ii) above is not always valid, the derived figure of merit remains valid for the design and optimization of experiments.

\subsection{Optimal placement of $\gamma$-ray detectors}

The optimal placement of $\gamma$-ray detectors for an IMPAC measurement may be determined by evaluating
\begin{equation}\label{eq:S2N}
S^2N \propto S^2(\theta_{\gamma}, d_{\gamma}) J_0(d_{\gamma}) W(\theta_{\gamma}, d_{\gamma}),
\end{equation}
where $\theta_{\gamma}$ is the angle with respect to the beam axis at which the detector is placed and $d_{\gamma}$ is the distance between the face of the detector crystal and the target. The slope $S$ and the angular correlation $W$ are functions of $d_{\gamma}$ through the solid-angle correction factors $Q_k$ for the $\gamma$-ray detectors. As defined by Rose \cite{ROSE.PhysRev.91.610}, the solid angle attenuation coefficients for a detector with cylindrical symmetry and its axis facing toward the beam spot, are given by
\begin{equation}
Q_k(E_{\gamma}) = J_k( E_{\gamma})/J_0( E_{\gamma}),
\end{equation}
where
\begin{equation}
J_k( E_{\gamma})=\int_0^{\beta_{\rm max}} P_k(\cos \beta) \varepsilon(E_{\gamma},\beta) \sin \beta d \beta.
\end{equation}
The integration variable $\beta$ is the angle between the direction of propagation of the $\gamma$ ray and the symmetry axis of the detector. The probability for photopeak detection along the direction $\beta$ for a $\gamma$ ray of energy $E_{\gamma}$ is $\varepsilon(E_{\gamma},\beta)$, which can be expressed approximately as
\begin{equation}
\varepsilon(E_{\gamma},\beta) = 1 - e^{-\mu(E_{\gamma})x(\beta)},
\end{equation}
where $x(\beta)$ represents the path length through the active region of the detector, and $\mu(E_{\gamma})$ is the energy-dependent $\gamma$-ray absorption coefficient. The absolute efficiency of the detector is given approximately by  $J_0( E_{\gamma})/2$ \cite{CAMP1969-192}.  The attenuation coefficients, $Q_k$, and $J_0$, were evaluated by the method of Krane \cite{KRANE1972-205,KRANE1973-401}, which gives values in close agreement with the more detailed calculations of Camp and Van Lehn \cite{CAMP1969-192}.

\begin{figure*}
\begin{center}
\resizebox{10cm}{!}{\includegraphics{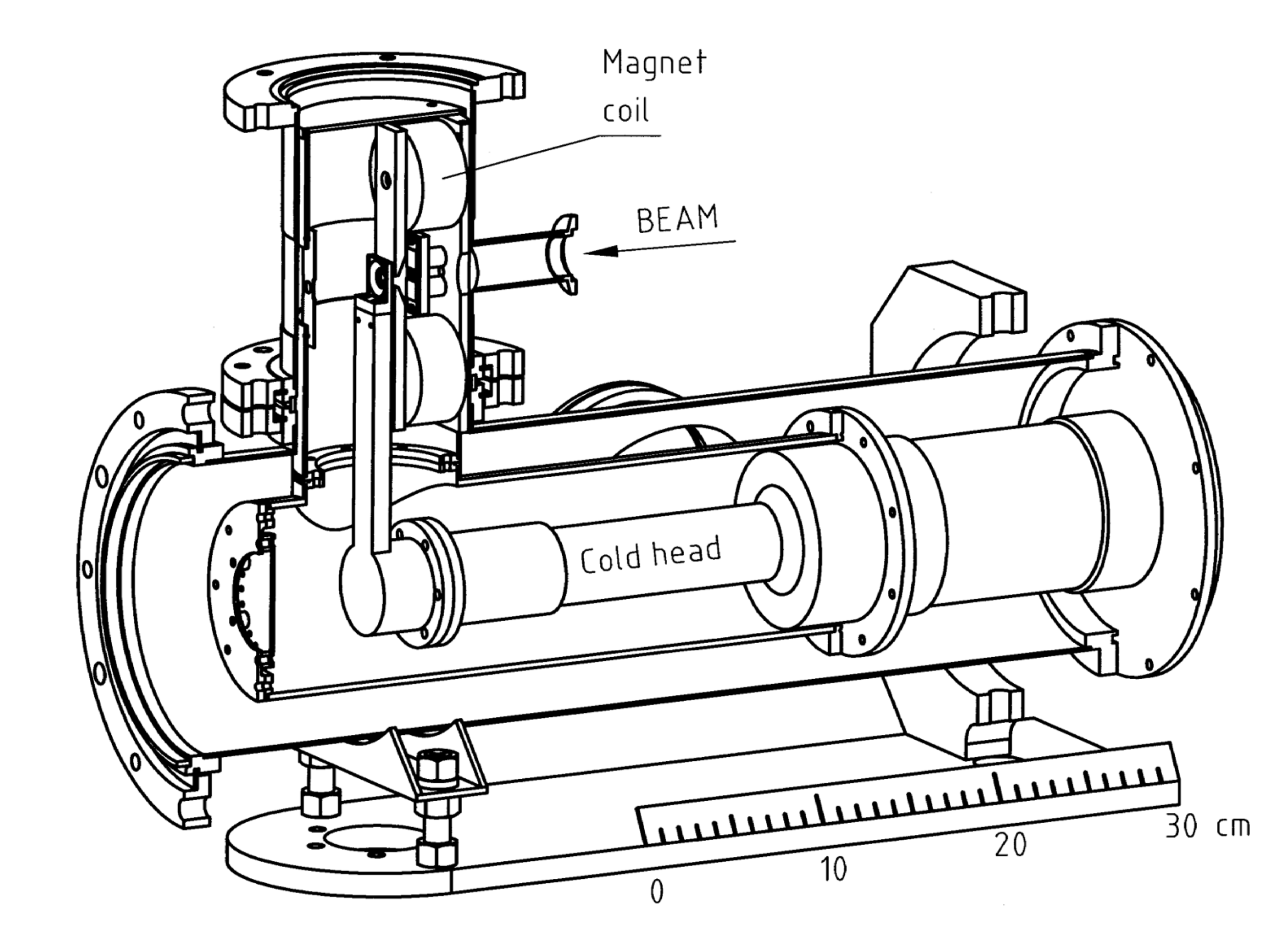}} \caption{Cut away view of the vacuum chamber. The cold head of the cryocooler, target support rod and magnet coils are visible. A pair of photodiodes which serve as particle detectors are placed between the magnet coils. A heat shield surrounding the cold head and target is attached to the first stage of the cryocooler. The electromagnet is attached to this heat shield.} \label{fig:coldhead}
\end{center}
\end{figure*}

\subsection{Examples of $\gamma$-ray detector placement}

Three examples of particle-$\gamma$ angular correlations following Coulomb excitation, with the detection of backscattered beam ions, are shown in Fig.~\ref{fig:s2N}, along with the corresponding sensitivity plots. The figure of merit for sensitivity scales with $\frac{1}{W}(\frac{dW}{d\theta})^2$. Thus the optimum placement angle of the $\gamma$-ray detectors tends to be on the smaller $W(\theta)$ side of the angle at which the angular correlation has its maximum slope.

These examples show that, on the whole, it is better to place the $\gamma$-ray detectors as close to the target as practical because the gain in count rate usually improves the sensitivity more than it is reduced by solid-angle attenuation, which reduces the anisotropy of the angular correlation. In real measurements it is rarely practical to place the detectors exactly at the angle of maximum sensitivity. For example, the Hyperfine Spectrometer is usually configured with the $\gamma$-ray detectors at $\pm 65^\circ$ to the beam axis to measure the precession effect via $E2$ transitions, although the optimum is closer to 70$^\circ$. The 65$^\circ$ placement reduces or avoids the effect of viewing the target through the target frame at 90$^\circ$ to the beam. For mixed multipolarity transitions the optimum angle is sometimes near 25$^\circ$ to the beam. In this case the detectors at $+25^\circ$ and $-25^\circ$ may touch. It has to be evaluated case-by-case whether it is better to increase the target-detector distance (pull the detectors back at $\pm25^\circ$) or shift the pair of detectors to a larger angle.

\subsection{Optimal particle detector placement and solid angle}

The optimal placement of the particle detector(s) in a transient-field IMPAC measurement cannot be evaluated in isolation from other experimental conditions such as the reaction kinematics and the target design. However the same rule of thumb that applies to $\gamma$-ray detectors still serves as a guide: Generally the gain in count rate from increasing the particle-detector solid angle wins out over the loss of anisotropy in the angular correlation.

To illustrate this point, consider a transient-field $g$-factor measurement aiming to measure $g(2^+_1)$ and $g(4^+_1)$ in $^{130}$Te following Coulomb excitation by a 195-MeV $^{58}$Ni beam \cite{Coombes2019a}. A useful figure of merit to optimize the particle detector geometry is
\begin{equation}
\label{eq:S2N-particle}
S^2 N \propto S^2W \langle \sigma x_{\rm tgt} \rangle
\end{equation}
where
\begin{equation}
\langle \sigma x_{\rm tgt} \rangle = \int_{x_{\rm tgt}} \int_{\theta_p} \int_{\phi_p}  \frac{{\rm d}^2 \sigma}{{\rm d} \Omega_p {\rm d} E} {\rm d} \Omega_p \frac{{\rm d}E}{dE/dx}
\end{equation}
is the Coulomb excitation cross section, $\frac{{\rm d}^2 \sigma}{{\rm d} \Omega_p {\rm d} E}$, integrated over the particle detector solid angle, ${\rm d}\Omega_p = \sin \theta_p {\rm d} \theta_p {\rm d} \phi_p$, and the target thickness, $x_{\rm tgt} $. $dE/dx$ is the stopping power of the beam in the target. $S^2W$ must be evaluated for the chosen $\gamma$-ray detection angle and based on statistical tensors averaged over the particle detector solid angle as described in \ref{sect:rectangular_tensors}.

The particle detector geometry adopted for the evaluation of $\gamma$-ray detector placement in Fig.~\ref{fig:s2N} serves as the standard for the following comparison. Specifically, a pair of rectangular particle detectors 3.8 mm above and below the beam axis, 10 mm high, 9 mm wide and 16 mm back from the target represents the standard configuration. Replacing these detectors by a pair that are 9 mm high and 25 mm wide increases $\langle \sigma x_{\rm tgt} \rangle$, and hence the count rate, by a factor of 2.3 while $S^2W$ decreases by a factor of 0.86, giving an overall gain of 2.0 in sensitivity for both the 2$^+_1$ and 4$^+_1$ states. In this comparison $S^2W$ was evaluated at $\theta_{\gamma}=65^{\circ}$. The increased particle detector solid angle must also be evaluated to check that the recoiling $^{130}$Te ions traverse the ferromagnetic layer of the target and stop in the non-magnetic backing layer (see Fig.~\ref{fig:target}).

\begin{figure}
\begin{center}
\resizebox{8cm}{!}{\includegraphics{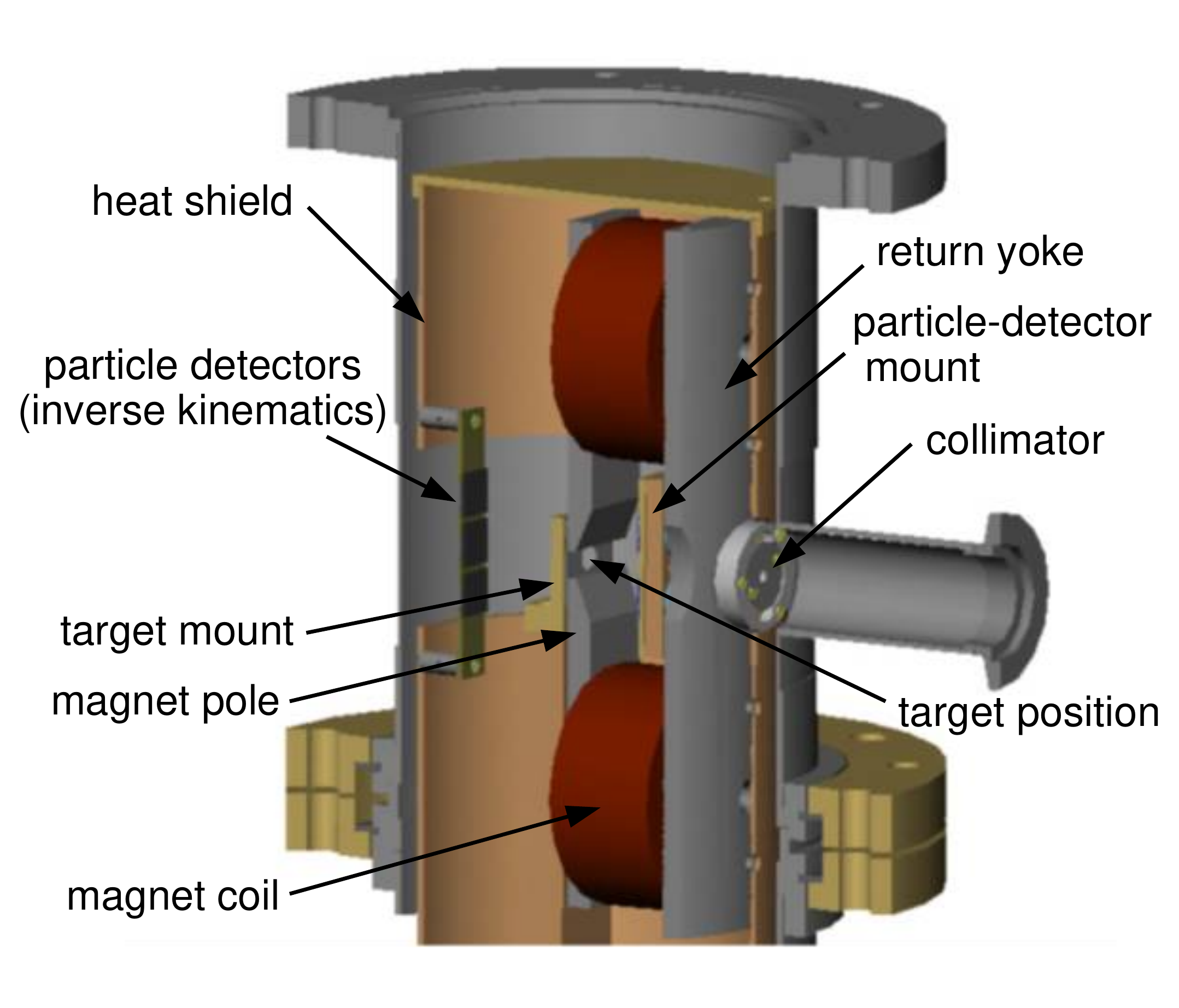}} \caption{Close up cut away view of the target chamber. The shaping of the magnet poles is shown as is the return yoke and collimator through which the beam enters. Note that the return yoke is shaped to remove material in the horizontal beam plane to reduce absorption of $\gamma$-rays. The three particle detectors at forward angles for measurements using reactions in inverse kinematics are indicated. (See also Fig.~\ref{fig:trifid} below.)}
\label{fig:closeup}
\end{center}
\end{figure}

\section{Apparatus design and description}
\label{sect:design}

\subsection{Overview and vacuum chamber}

A cross-sectional view of the target chamber and cryocooler, with the magnet and target mount in place, is presented in Fig.~\ref{fig:coldhead}. The view is looking toward the beam direction obliquely from an angle of about +45$^{\circ}$ to the beam axis. The target chamber is located above the cryocooler with the target attached to a vertical mounting rod that is attached to the second stage of the cold head. The electromagnet is attached, via its return yoke, to an inner shroud that serves as a support and heat shield, which is attached to the first stage of the cryocooler. To minimize $\gamma$-ray absorption in the horizontal plane through the beam axis, the shroud is milled out and there covered with aluminium foil. Beam enters through a hole in the return yoke. Particle detectors are located between the magnet coils, above and below the beam axis, to detect backscattered beam ions. A Pb foil is mounted on (but insulated from) the heat shield and located downstream of the target to serve as a beam stop.

Figure~\ref{fig:closeup} shows a closeup view of the vacuum chamber, generated from the CAD (Computer Aided Design and Drafting) drawing for the project. The view is looking (obliquely) from the beam direction. This figure indicates the collimator mounted onto the return yoke of the electromagnet through which the beam passes. Current can be read on the collimator. It has a 3~mm diameter opening and is located 46~mm from the target. Tantalum and molybdenum collimators are available.
Moving along the beam direction towards the target position, the shaping of the pole tips is evident. (The magnetic field will be discussed in more detail below.) This image does not show the vertical extension of the cold head but it does include the target mount that detaches from it. Three particle detectors are indicated downstream of the target; this arrangement is typical for a transient-field $g$-factor measurement in inverse kinematics.

\begin{figure*}[t]
\begin{center}
\resizebox{9.5cm}{!}{\includegraphics{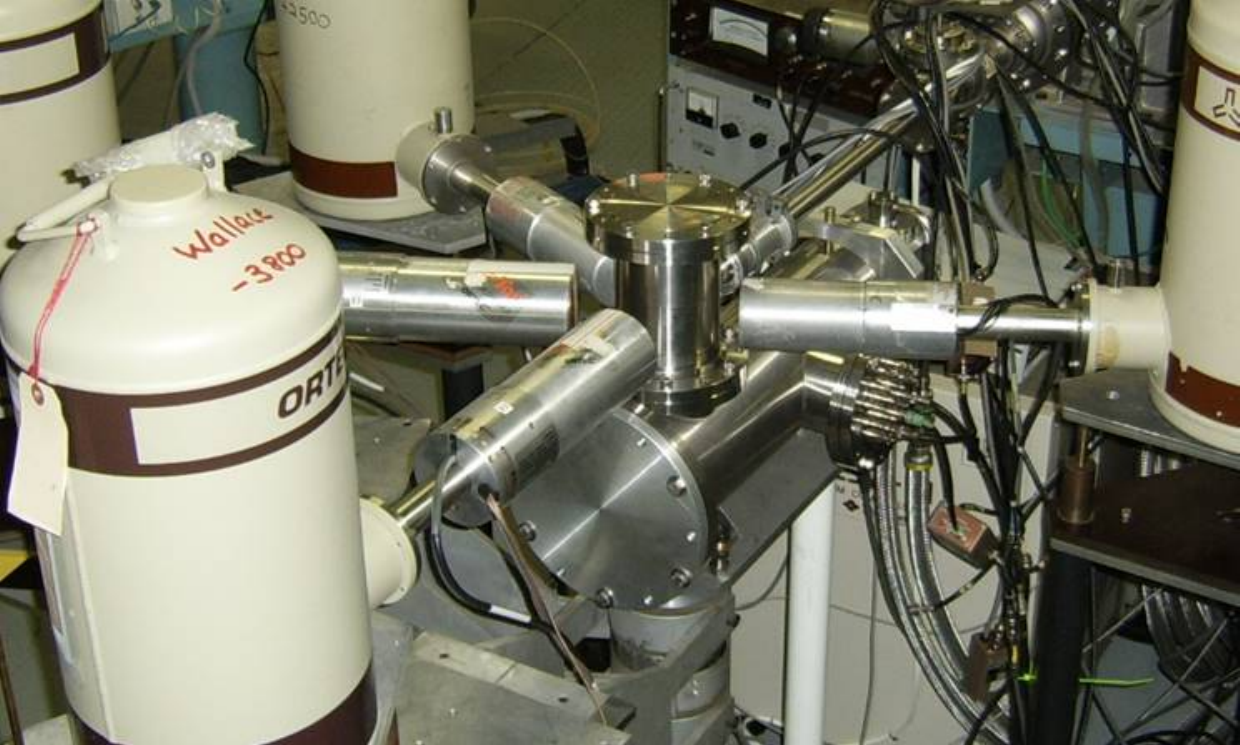}} \caption{The hyperfine spectrometer on the beamline, instrumented with four HPGe detectors.} \label{fig:lab-photo}
\end{center}
\end{figure*}

Figure~\ref{fig:lab-photo} shows the apparatus on the beam line with four HPGe detectors in place for an angular correlation measurement. The Hyperfine Spectrometer was installed on a pre-existing angular correlation table that provides rotating supports for four detectors. The angular scale is marked in 1$^{\circ}$ steps. While the detector can be aligned accurately with respect to the angular correlation table, the usual procedure we have used to check the detector position is to measure the angular correlation for at least the pair of detectors at forward angles. A comparison of the experimental and theoretical angular correlations will expose an offset in the set angle, if present, at the level of about one degree. This procedure has the advantage that it takes account of any displacement of the beam spot from the centre of the target, and likewise any displacement of the detector crystal from the centre of its vacuum capsule.

\subsection{Cryocooler and temperature control}

\begin{figure}[h]
\begin{center}
\resizebox{8cm}{!}{\includegraphics{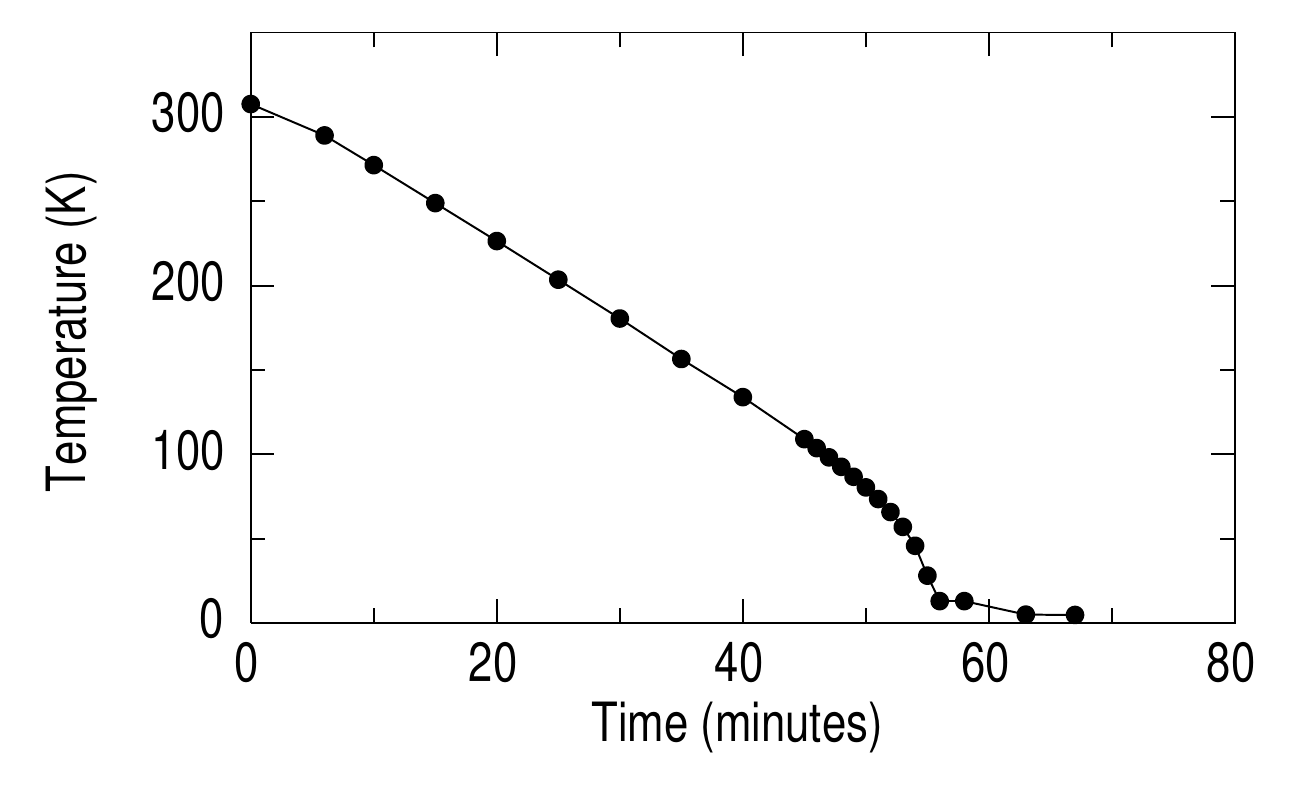}} \caption{Target temperature as a function of time from cryocooler power on.} \label{fig:cooldown}
\end{center}
\end{figure}

The cryocooler is a Sumitomo model RDK 408D with a cooling power of 1 W at 4~K  and about 10 W at 10~K. A LakeShore Model DT-670C-SD temperature sensor is located just below the removable target mount. Figure~\ref{fig:cooldown} shows the target temperature as a function of time during cool down. Without heat load the target temperature can reach below 3~K, however with the target in place and under beam, the monitored temperature is typically $\sim 5$~K. Evidence will be presented below that the temperature within the beam spot may be some 10 to 20 degrees warmer, depending on the beam power deposited in the target.

Temperature control is achieved by use of a LakeShore Controller Model 331, which uses a PID (Proportional, Integral, Derivative) feedback loop to adjust the current through a LakeShore 50 W heater that is attached to the lower part of the vertical target rod. The monitored temperature of the target can be held constant to within a fraction of a degree under steady beam conditions. To date only a few measurements have been performed where target temperature is critical.

\begin{figure}[h]
\begin{center}
\resizebox{8cm}{!}{\includegraphics{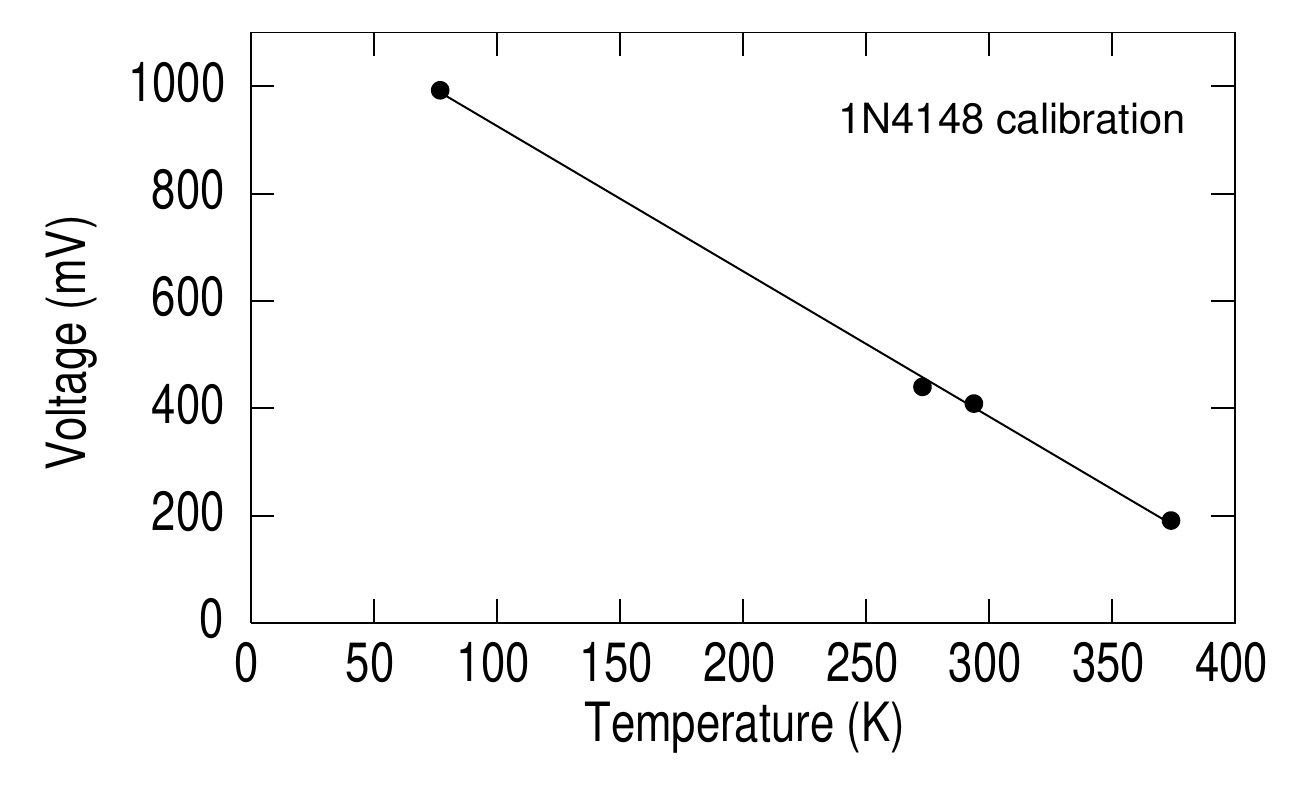}} \caption{Calibration of a 1N4148 diode as a temperature sensor to be used with the LakeShore Model 331 controller.} \label{fig:stage1temp}
\end{center}
\end{figure}

To assist warm-up, a 50~W 50 ohm aluminium-housed wire-wound resistor is attached to the first stage (on the flange visible to the left of the cold head in Fig.~\ref{fig:coldhead}). An ordinary small signal switching diode 1N4148 is used to monitor the first-stage temperature. It connects directly to the LakeShore Controller input and was calibrated at four temperatures between 77~K and 374~K. As shown in Fig.~\ref{fig:stage1temp}, the 1N4148 response (i.e. the voltage drop across the temperature-dependent resistance) is a linear function of temperature over this range. During operation the stage-1 temperature stabilizes between 90 and 100~K.

For optimal cooling, the target mount is made of copper and connected directly to the second stage of the cryocooler. The copper surfaces are nickel plated to prevent oxidation. As a consequence of the all metal connection between the target and the cryocooler, the beam current on target cannot be measured in the conventional way, at least for thick targets that stop the beam. For targets that transmit the beam, current can be read on the beam stop in front of the target.

In cases where it is not possible to measure the target current, it has proved equally as effective to monitor the beam intensity via the count rate in the particle detectors that register beam ions scattered from the target. The analog output of an Ortec model 661 rate meter can be fed to the accelerator beam current metering system to assist with beam tuning.

\subsection{Magnet Design}

The requirements for the electromagnet design were that (i) it must be compact, (ii) the field applied to the target should be at least 0.07 T for targets with iron as the host \cite{Stuch98NMM}, and of order 0.1 T for targets with gadolinium as the host \cite{HAUSSER1983,ROBINSON1999}, (iii) beam bending must be minimized.

\begin{figure}
\begin{center}
\resizebox{7cm}{!}{\includegraphics{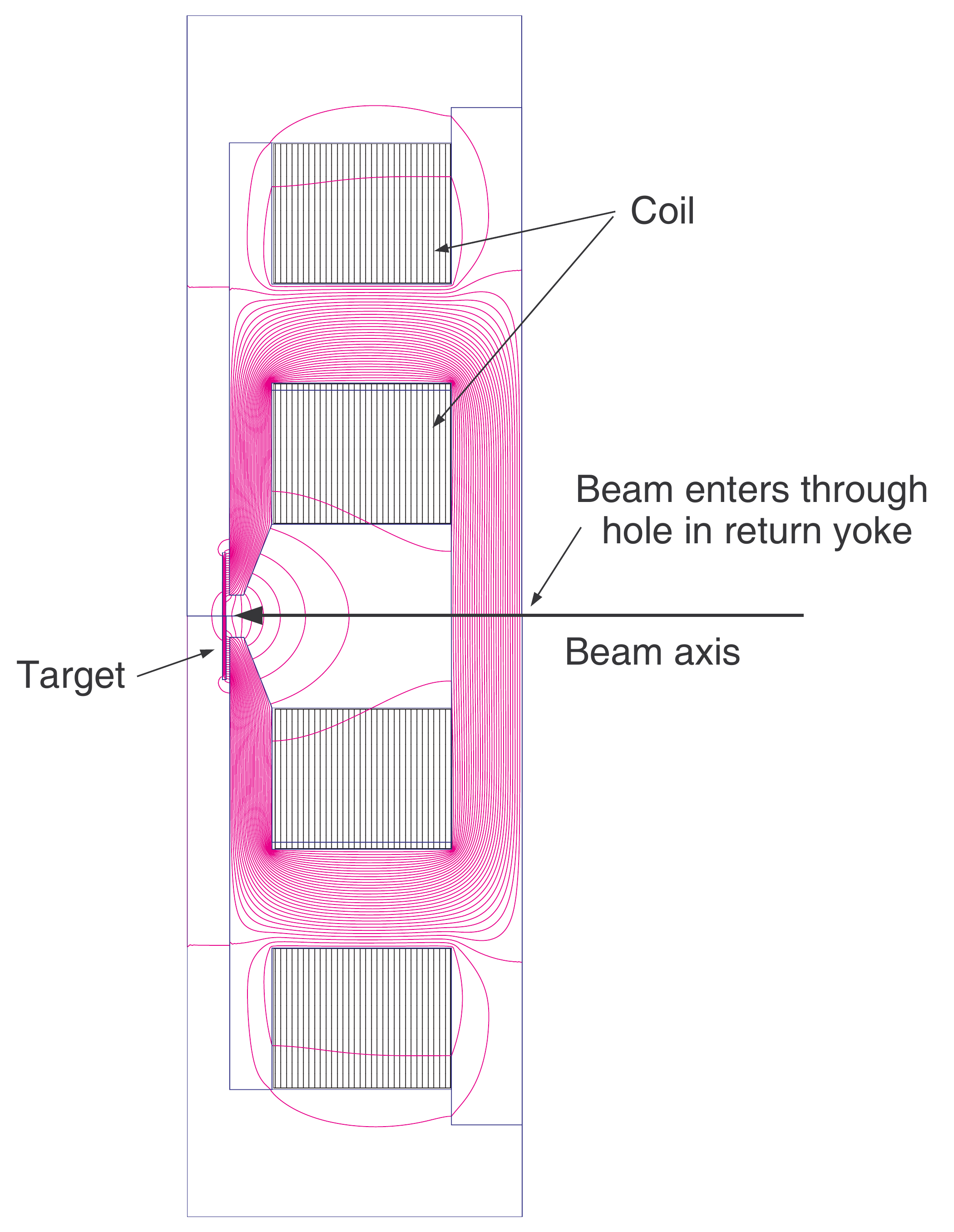}} \caption{Poisson Superfish \cite{halbach1976superfish} calculations of the magnetic field produced by the electromagnet. The pole gap and pole-piece shape in this final design was chosen after a systematic study of alternative pole geometries to maximize the field in the target location while minimizing the stray field elsewhere (particularly along the beam path).}
\label{fig:magnet-circuit}
\end{center}
\end{figure}

The present design was chosen on the basis of previous experience with several electromagnet designs and pole-piece geometries \cite{MPRthesis,AESthesis,CGRthesis,CEDthesis}, along with detailed calculations using the {\em Poisson Superfish} software \cite{halbach1976superfish}. The {\em Superfish} calculation for the final design is shown in Fig.~\ref{fig:magnet-circuit}. A feature of the present design is that the use of soft iron cones, which were included in previous apparatus (in our lab \cite{AESthesis,CGRthesis,CEDthesis} and elsewhere \cite{Holthuizenthesis,LESKE2003_Ne20}) to shield the incident beam path from the stray magnetic field, were omitted. The {\em Superfish} calculations indicated that the shielding cones pulled flux away from the pole gap, hence reducing the polarizing field applied to the target as well as extending the stray field along the beam path. The decision was therefore taken to remove them altogether. Apparatus constructed more recently for transient-field experiments on radioactive beams at ISOLDE \cite{ILLANA2015,Zn72_LVTF_PhysRevC.89.054316} used the same approach.

\begin{figure}
\begin{center}
\resizebox{8cm}{!}{\includegraphics{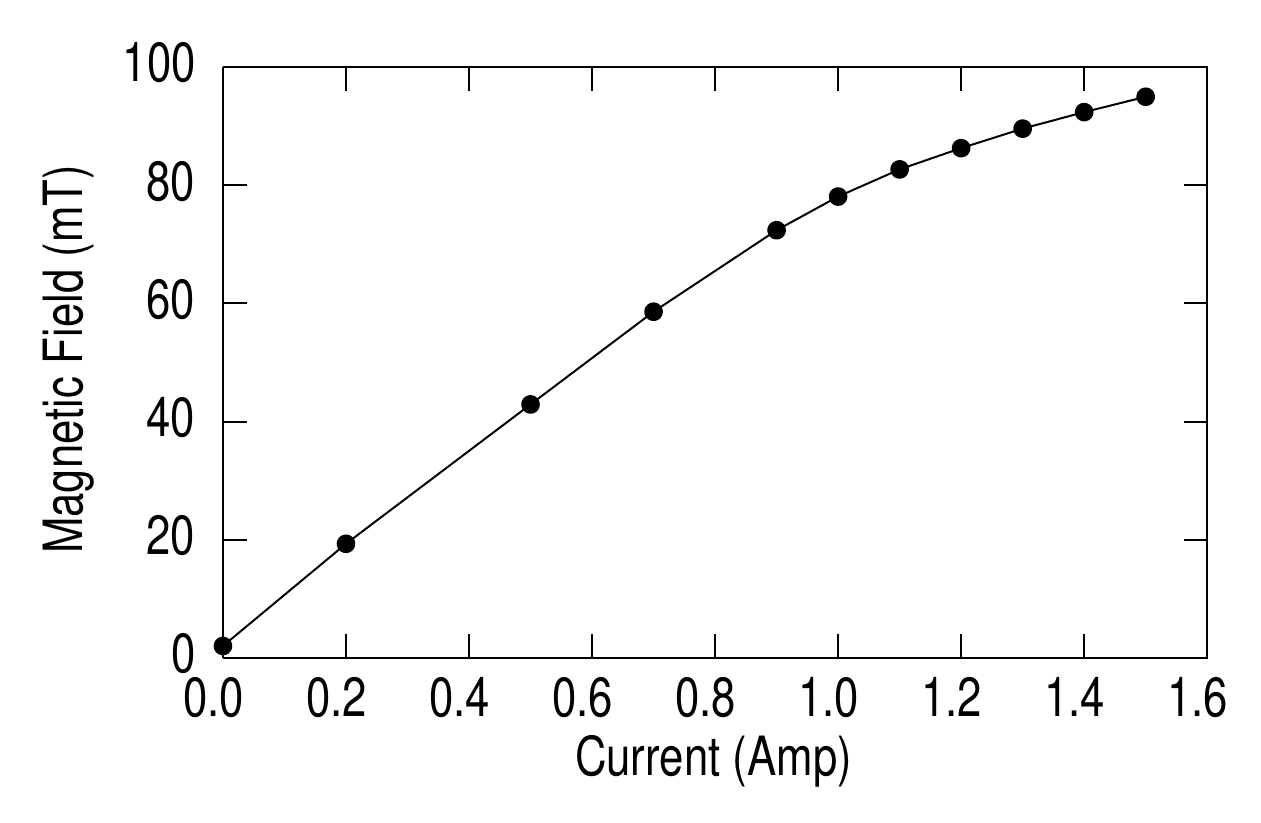}} \caption{Measured magnetic field at the target location (with the target removed) as a function of current through the coils.} \label{fig:BvsI}
\end{center}
\end{figure}

Calculations showed that the bending of the heavy-ion beams along the beam path in typical experiments was negligible, in line with the experience amassed in our laboratory and elsewhere \cite{ben80,speidel02,ILLANA2015}.

Figure~\ref{fig:BvsI} shows the measured magnetic field between the pole tips versus the current through the coils. The {\em Superfish} calculations predicted saturation of the target at 0.07-0.08 T, which requires a coil current of $\sim 1$ A. Because these magnetic field calculations were performed in two-dimensions using Cartesian geometry, the calculated values of magnetic field are slightly overestimated. A slightly smaller field in the pole gap was observed than calculated. In practice, the standard procedure is to operate the magnet with a current of 1.3~A thus producing a measured field of 0.09 T at the target location with the target removed.

The magnet assembly is mounted on the first stage of the cryocooler and typically equilibrates at a temperature of $\sim 90$~K. Cooling the magnet has the advantage of reducing the resistance of the coils and thus the heat dissipated. However, the disadvantage is that while the system is under vacuum there is a very poor thermal path between the magnet coils and the ``outside world''. Thus the cryocooler must always be operated while the magnet is energized under vacuum to remove heat and prevent the coils from overheating.

\subsection{Particle detection}

\begin{figure}
\begin{center}
\resizebox{7cm}{!}{\includegraphics{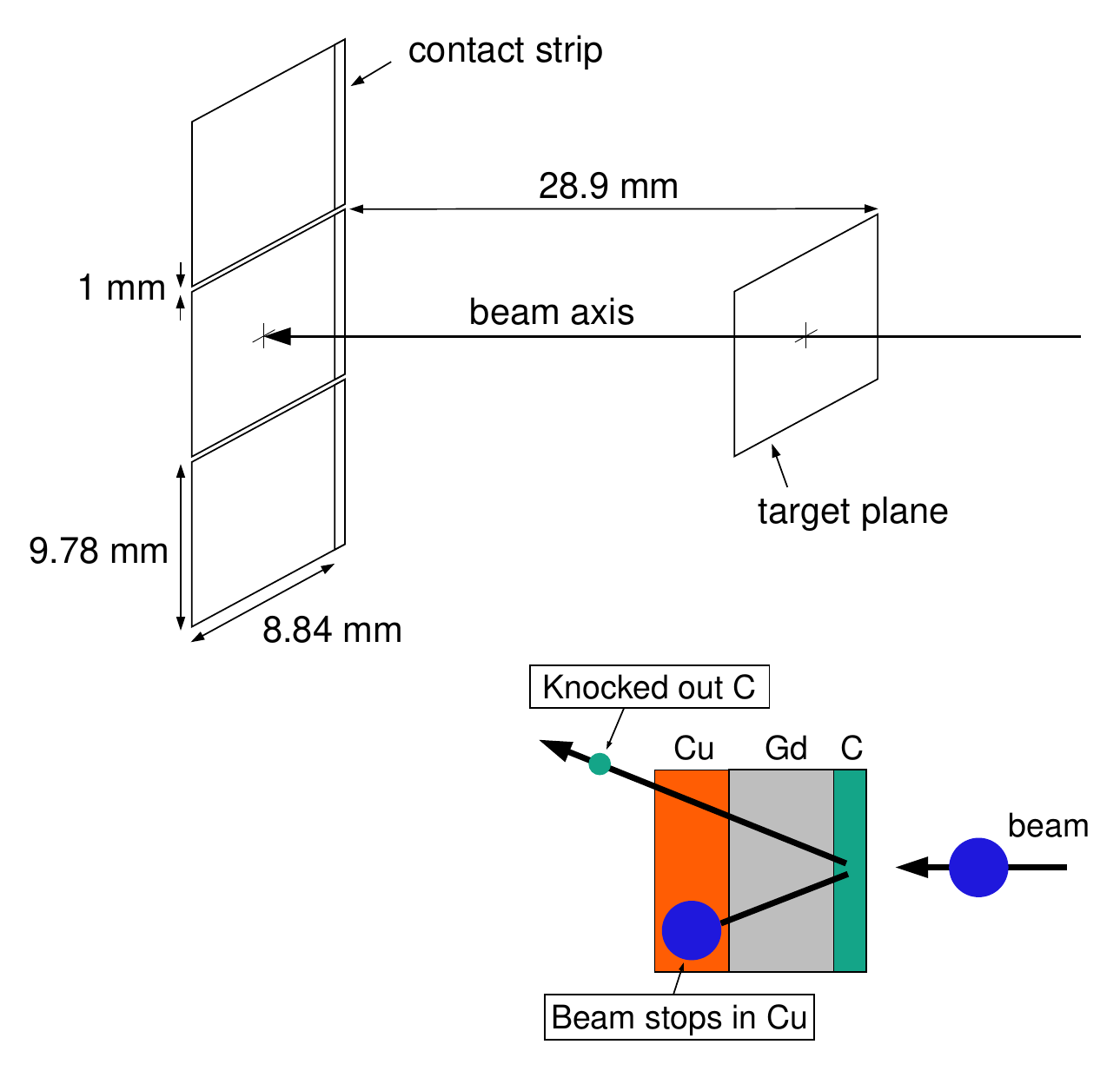}} \caption{Schematic view of the arrangement of the particle detectors at forward angles as used for inverse reaction kinematics. The dimensions of the active area are indicated on the lower detector. The graphic represents a three-layer target for a transient-field $g$-factor measurement on excited beam nuclei; the heavier beam ion knocks forward a C target ion which is detected at forward angles. The scattered beam ion traverses the magnetized gadolinium layer of the target and stops in a copper backing.}
\label{fig:trifid}
\end{center}
\end{figure}

To date particle detection has been achieved with silicon photodiodes as heavy-ion particle detectors. For the most part these have been solderable photodiodes from Advanced Photonix Inc., model number PDB-C613-2, with active area 9.78 mm $\times$ 8.84 mm (and contact strip along the 9.78 mm side). Figure \ref{fig:trifid} shows the arrangement of three of these diodes for measurements in inverse kinematics. The arrangement for conventional (backscatter) kinematics places these detectors 16.2~mm from the target and 4 mm above and below the beam axis. In this case the detectors are oriented such that the contact strip is at the top(bottom) of the the top(bottom) detector. To increase the count rate in transient-field $g$-factor measurements, it has been found feasible to use PDB-C615-2 photodiodes at the backward location placed so as to give a sensitive vertical height of 9.25~mm and a width of 25.17~mm.

While it is possible to connect the photodiodes in parallel before feeding the signals into a suitable preamplifier, they generally have been operated with separate electronics channels, which gives additional information for diagnostic purposes at the cost of some additional effort to analyze the data.

\subsection{Electronics}

The electronics setup depends on the type of measurement and is reconfigured using dedicated modules as required. Rather than attempt a comprehensive description, the  analog electronics for IMPAC experiments will be outlined in this section to illustrate the main features and describe the procedures associated with reversing the direction of the magnetic field and recording the field direction in the data stream. (The Hyperfine Spectrometer project also included building a second general purpose data acquisition system for the Heavy Ion Accelerator Facility that will not be described in any detail here.)

\begin{figure}
\begin{center}
\resizebox{8cm}{!}{\includegraphics{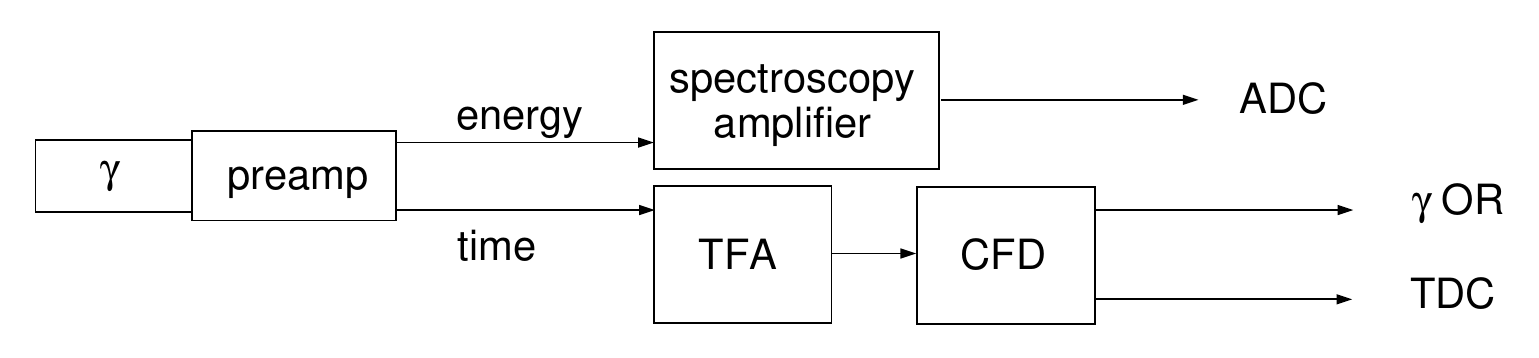}}
\caption{Electronics for $\gamma$-ray detectors (one of four) showing the complete path to the ADC for the energy signal and the front end of the fast-timing and trigger-logic path.}
\label{fig:gamma-electronics}
\end{center}
\end{figure}

Figure~\ref{fig:gamma-electronics} shows the basic electronics associated with each $\gamma$-ray detector. This is a standard arrangement. There are two signal streams - one for the energy signal and one for timing, which originate from the two identical output signals from the HPGe detector preamplifier. The energy signal is passed directly to a spectroscopy amplifier (ORTEC 572) and then to an analog to digital converter (ADC) in the data acquisition system (DAQ). The timing signal passes through a timing filter amplifier (TFA; ORTEC 474) to a Constant Fraction Discriminator (CFD: ORTEC 473) which provides timing logic signals. One output is fed directly to a time to digital converter (TDC) in the data acquisition system to generate particle-$\gamma$ time differences, while the other is used to generate the master gate (or DAQ trigger) - see below.  The TDC is a 16 channel CAEN  module V1290 N.

\begin{figure}
\begin{center}
\resizebox{8cm}{!}{\includegraphics{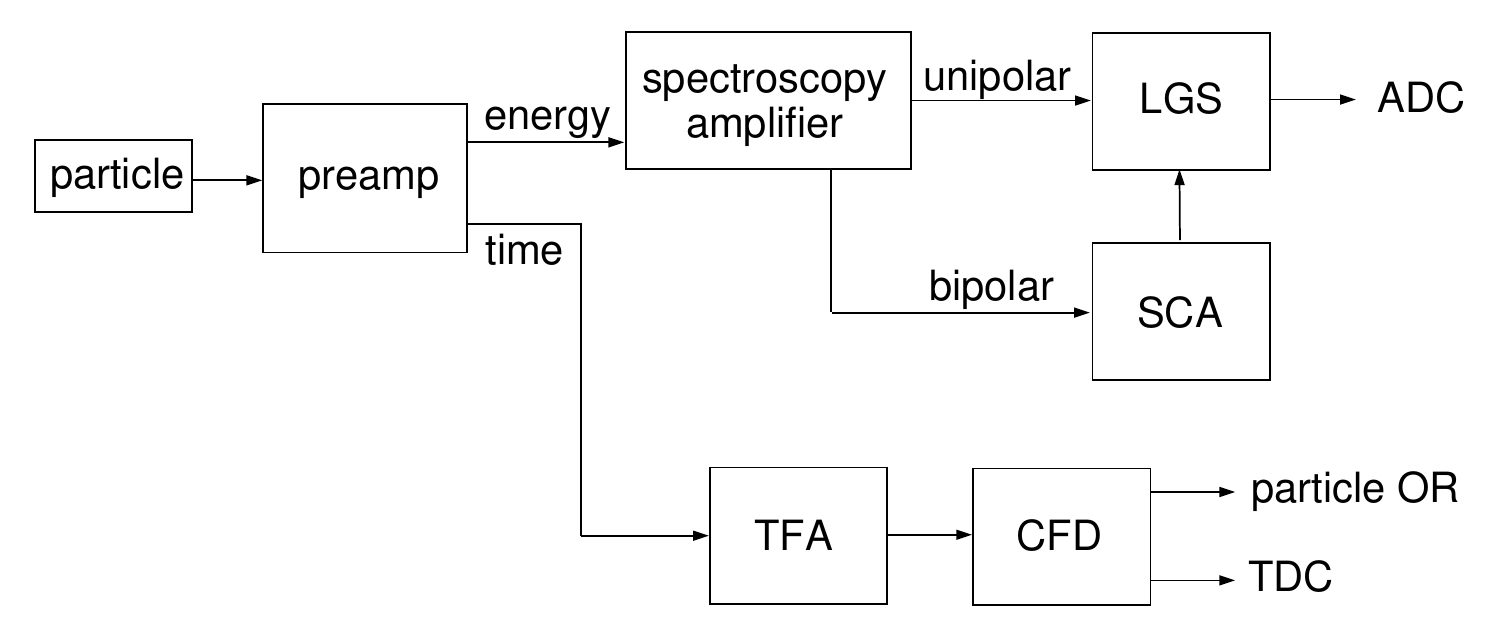}}
\caption{Electronics for particle detectors (one of two) showing the complete path to the ADC for the energy signal and the front end of the fast-timing and trigger-logic path. }
\label{fig:particle-electronics}
\end{center}
\end{figure}

As shown in Fig.~\ref{fig:particle-electronics}, the particle detector electronics is conceptually similar to that of the $\gamma$ detectors. One difference is that the timing output of the particle preamplifier (ORTEC 142B) is shaped to drive a CFD, but this feature cannot be used with the photodiode detectors as the signals have the wrong polarity. It is possible to modify the preamplifiers to reverse the polarity, but we opted to use a TFA with no shaping to invert the polarity for input the the CFD. The energy signal is again fed directly to a spectroscopy amplifier; however in this case a linear gate and stretcher (LGS; ORTEC 542) gated by a single channel analyzer (SCA) is included to cut out low energy noise before the signal is presented to the ADC.

\begin{figure}
\begin{center}
\resizebox{8cm}{!}{\includegraphics{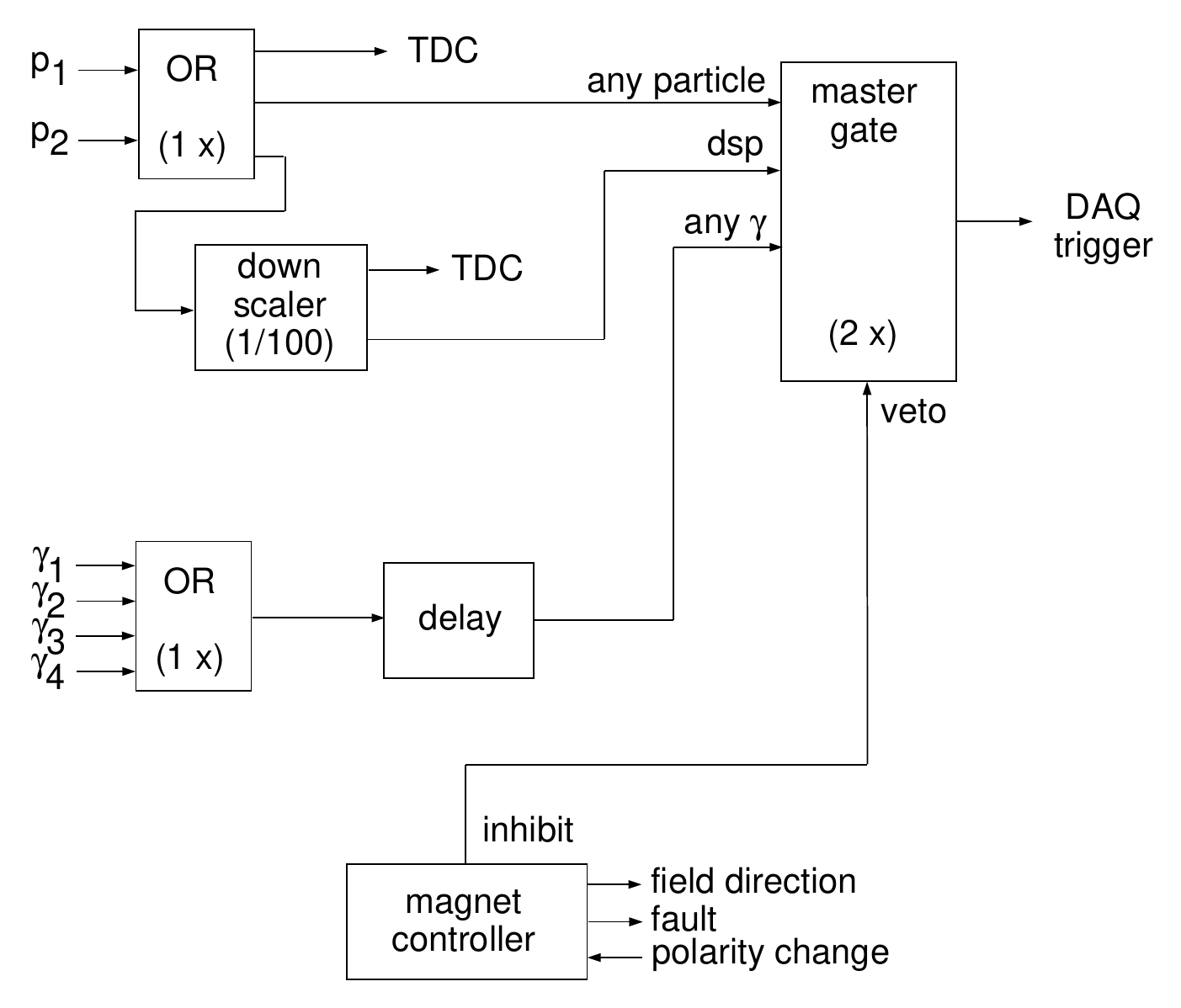}}
\caption{Electronics for generating the master gate or DAQ trigger. The data acquisition is triggered when there is a particle-$\gamma$ coincidence or if there is a down-scaled particle (dsp) singles event.}
\label{fig:trigger-logic}
\end{center}
\end{figure}

Figure \ref{fig:trigger-logic} represents the logic modules used to generate the master gate (or DAQ trigger), based largely on Phillips Scientific model 755 Quad Majority logic units. Timing logic pulses are generated for `any particle' (an OR of the two particle CFD outputs) and for `any gamma'  (an OR of the four $\gamma$-detector CFD outputs). The basic coincidence requirement is the detection of a particle-$\gamma$ coincidence between either of the two particle detectors and any one of the four $\gamma$-ray detectors. The particle-$\gamma$ trigger is therefore generated by overlapping (forming an AND of) the `any particle' and `any gamma' logic signals. To this end, the width of the `any particle' logic pulse is set to 400 ns whereas the `any gamma' pulse is delayed by approximately 200 ns (by adjusting the time-width of the $\overline{OUT}$ signal to 200 ns and triggering another channel of the logic unit on the falling edge). Thus particle-$\gamma$ events in the TDC are recorded in a time window of $\pm 400$~ns around the prompt-coincidence peak.

In addition to the particle-$\gamma$ coincidence, it is useful to record down-scaled particle-singles events in the data stream. A copy of the `any particle' pulse train is down scaled (typically by a factor of 100). One copy of the output of the down-scaler is sent to the TDC to label the event. The other copy is presented to the master gate module set on a coincidence requirement of 2: the `any particle' pulse is always present when there is a down-scaled particle event. Thus the DAQ will be triggered for both particle-$\gamma$ and down-scaled particle events, as well as the occasional events where the down-scaled particle is also in coincidence with a $\gamma$ ray.

\begin{figure}
\begin{center}
\resizebox{8cm}{!}{\includegraphics{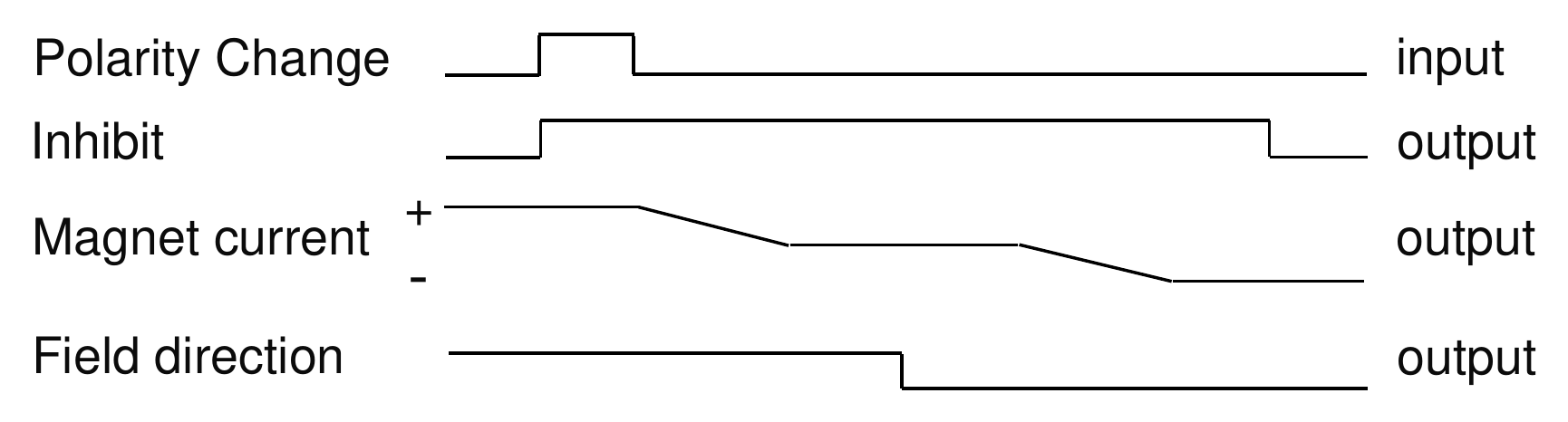}}
\caption{Timing sequence for magnet control unit during a change in magnetic field direction. The polarity change, inhibit and field direction signals are logic pulses or levels.}
\label{fig:TimingSeq}
\end{center}
\end{figure}

Finally, the master gate module is vetoed by the magnet controller (see below) whenever the magnetic field is in the process of being reversed, or if the magnet controller detects a fault condition (e.g. no current through coils).

The magnet is energized by a programmable constant-current power supply which is controlled by a custom module. Figure \ref{fig:TimingSeq} shows the time sequence for the input and outputs of the magnet controller during a field flip. The field direction is fed into the DAQ via a `user word' that mimics an ADC. The adopted convention has been to label field up(down) events with channel number 64(8) in the field-direction spectrum.

During the measurements, the direction of the magnetic field is reversed periodically (typically every 15 minutes). For this purpose a recycling scaler counting particle singles sends an output pulse to trigger the polarity change when the scaler reaches its preset limit, which the experimenter can set as desired.

\section{Apparatus performance}
\label{sect:performance}

\subsection{Cool down and warm up}

The cool down time for the target is about an hour, as shown in Fig.~\ref{fig:cooldown}. The temperature of the heat shield connected to the first stage of the cryocooler takes a few hours to stabilize (somewhat below 100~K). Warm-up is achieved by driving 50~W heating elements connected to the first and second stages via the LakeShore controller with a set temperature of $\sim 300$~K. Warm-up typically takes a few hours, largely due to the thermal mass of the heat shield.

\begin{figure}
\begin{center}
\resizebox{8cm}{!}{\includegraphics{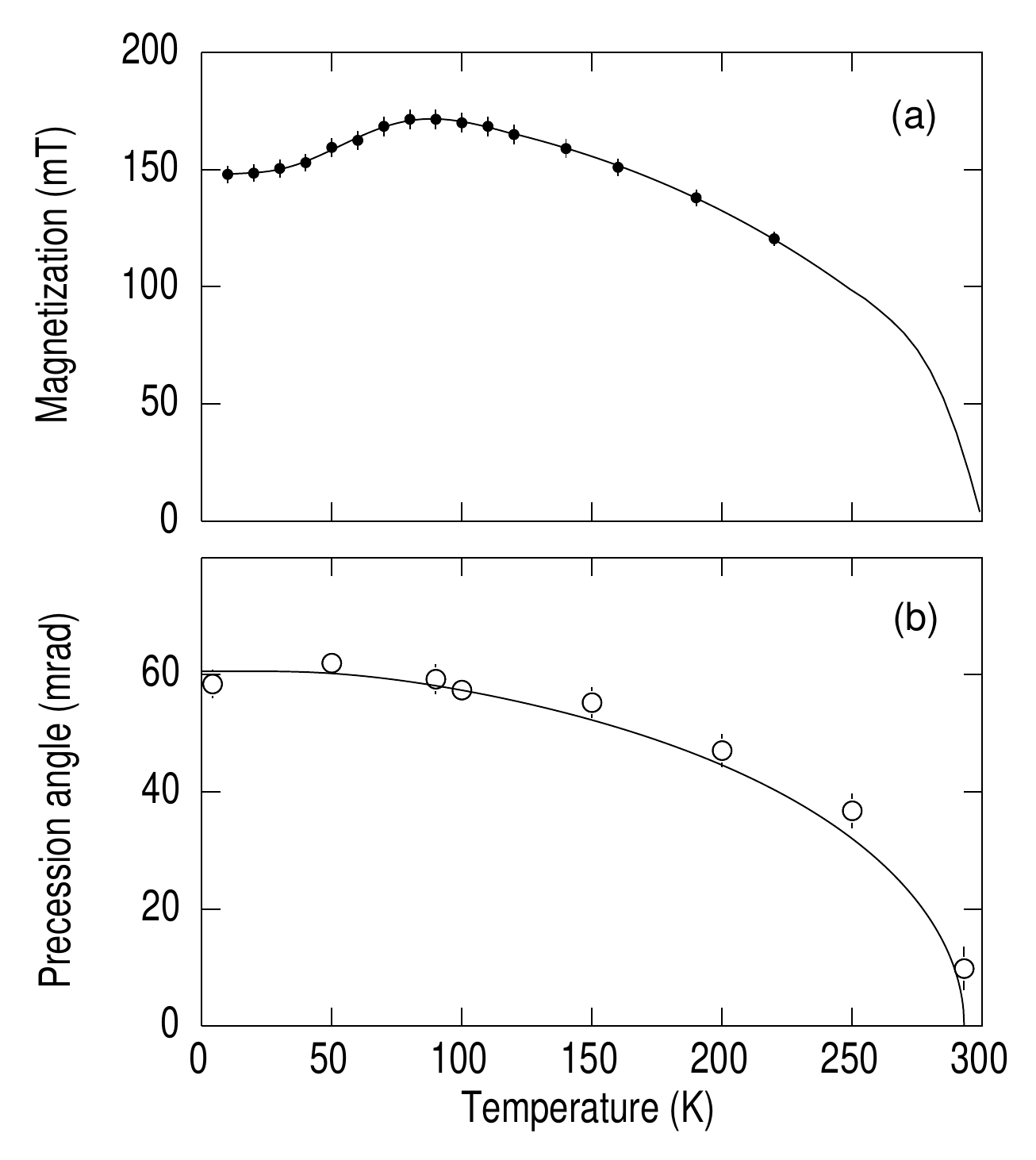}}
\caption{(a) Magnetization versus temperature for a typical rolled and annealed gadolinium foil. The points are measurements with the Rutgers magnetometer \cite{PIQUE1989}. The line to guide the eye above 220~K is informed by magnetization measurements reported in Refs.\cite{Gd_mag_PhysRev.132.1092} and \cite{Darnell64}. (b) Measurements of transient-field precession angles for $^{196}$Pt in a gadolinium foil as a function of temperature. The solid line represents the Brillouin function for $T_C=293$~K.}
\label{fig:magGd}
\end{center}
\end{figure}

\begin{figure}
\begin{center}
\resizebox{8cm}{!}{\includegraphics{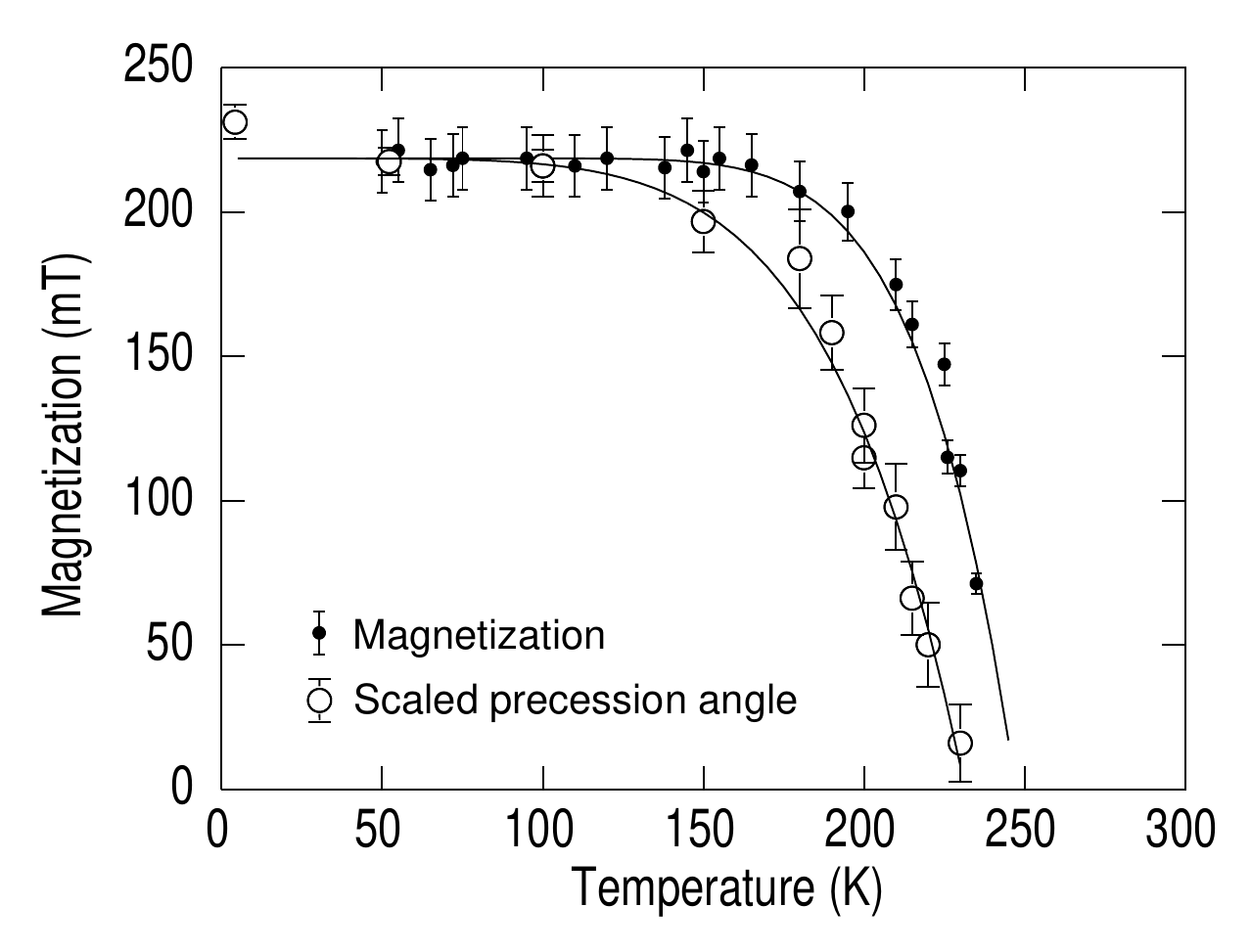}} \caption{Magnetization of a terbium foil as a function of temperature measured with the Rutgers magnetometer \cite{PIQUE1989} compared with scaled transient-field precession angles. The fall off in the precession angle compared to the magnetization as the Curie temperature of 222~K is approached is interpreted as evidence of an increased temperature in the beam spot.}
\label{fig:magnetizn}
\end{center}
\end{figure}

\subsection{Beam heating}

Figure \ref{fig:magGd}(a) shows a typical magnetization curve for a gadolinium foil $\sim 5$~mg/cm$^2$ thick polarized by an external field of 0.08~T. These measurements were made with the Rutgers magnetometer \cite{PIQUE1989}. The dip at low temperature indicates that the rolled and annealed foil has its microcrystals aligned so that the basal planes are in the plane of the foil (i.e. $b$-axis is in direction of the applied magnetic field and the $c$-axis is perpendicular to the plane of the foil, i.e. along the beam direction) \cite{Gd_mag_PhysRev.132.1092}. Figure~\ref{fig:magGd}(b) shows the temperature dependence of the transient-field precession angles measured for $^{196}$Pt traversing a similar gadolinium foil using the Hyperfine Spectrometer. The polarizing field was 0.09~T. The low-temperature dip in the magnetization is not clearly evident in the precession angles; there may be a hint of a reduced field at 5~K but the dip in the magnetization is washed out. Otherwise, the precession angles largely follow the magnetization. The Curie temperature of gadolinium is 293~K \cite{Gd_mag_PhysRev.132.1092}. Figure~\ref{fig:magnetizn} compares the measured magnetization together with scaled transient-field precession angles for $^{196}$Pt traversing a terbium foil as a function of temperature. Again, the measurements were made using the Rutgers magnetometer \cite{PIQUE1989} and the Hyperfine Spectrometer. The Curie temperature of terbium is 222~K.

In both Fig.~\ref{fig:magGd} and Fig.~\ref{fig:magnetizn}, the $^{196}$Pt nuclei were Coulomb excited by a 170-MeV $^{58}$Ni beam, with a beam current of $\sim 1$~pnA. Transient-field precessions are expected to follow the magnetization of the host. In the case of the gadolinium host, the pronounced dip in magnetization at low temperature was not observed in the transient-field precession data; for the terbium host it is very clear that the measured transient-field precessions begin to decrease about 20~K below the Curie temperature of the host. The conclusion is that the average temperature within the beam spot under these conditions could be $\sim 20$~K higher than the temperature read on the target mount. In fact, for a set target temperature of 150~K, somewhat below the Curie temperature of terbium, the transient-field  precession angle has been observed to decrease as the beam intensity increases. Fortunately the measured precession angles for both hosts become essentially constant for temperatures below $\sim 100$~K. It is in this regime, where a temperature rise of some tens of degrees K has little effect on the observed precession angle, that the transient-field measurements employing gadolinium hosts have been performed over the past decades. (See the review articles from the Bonn \cite{speidel02} and Rutgers \cite{benczerkoller07} groups. Other examples include Refs.~\cite{East.PhysRevC.79.024303,Chamoli.PhysRevC.83.054318,STUCHBERY2006.magnetization,Zn72_PhysRevC.85.034334,Walker.PhysRevC.84.014319}.)

Thus beam heating is not a major concern for transient-field $g$-factor measurements where modest beam currents of the order of a few pnA are used. Our strategy is generally to run the cryocooler with no temperature feedback during $g$-factor measurements, which gives a temperature on the target mount of $\sim 5$~K. However if the objective is to study the temperature dependence of hyperfine interaction effects, like those in Fig.~\ref{fig:magGd} and Fig.~\ref{fig:magnetizn}, then care must be taken to minimize beam heating and evaluate its effects.


\begin{figure}
\begin{center}
\resizebox{7cm}{!}{\includegraphics{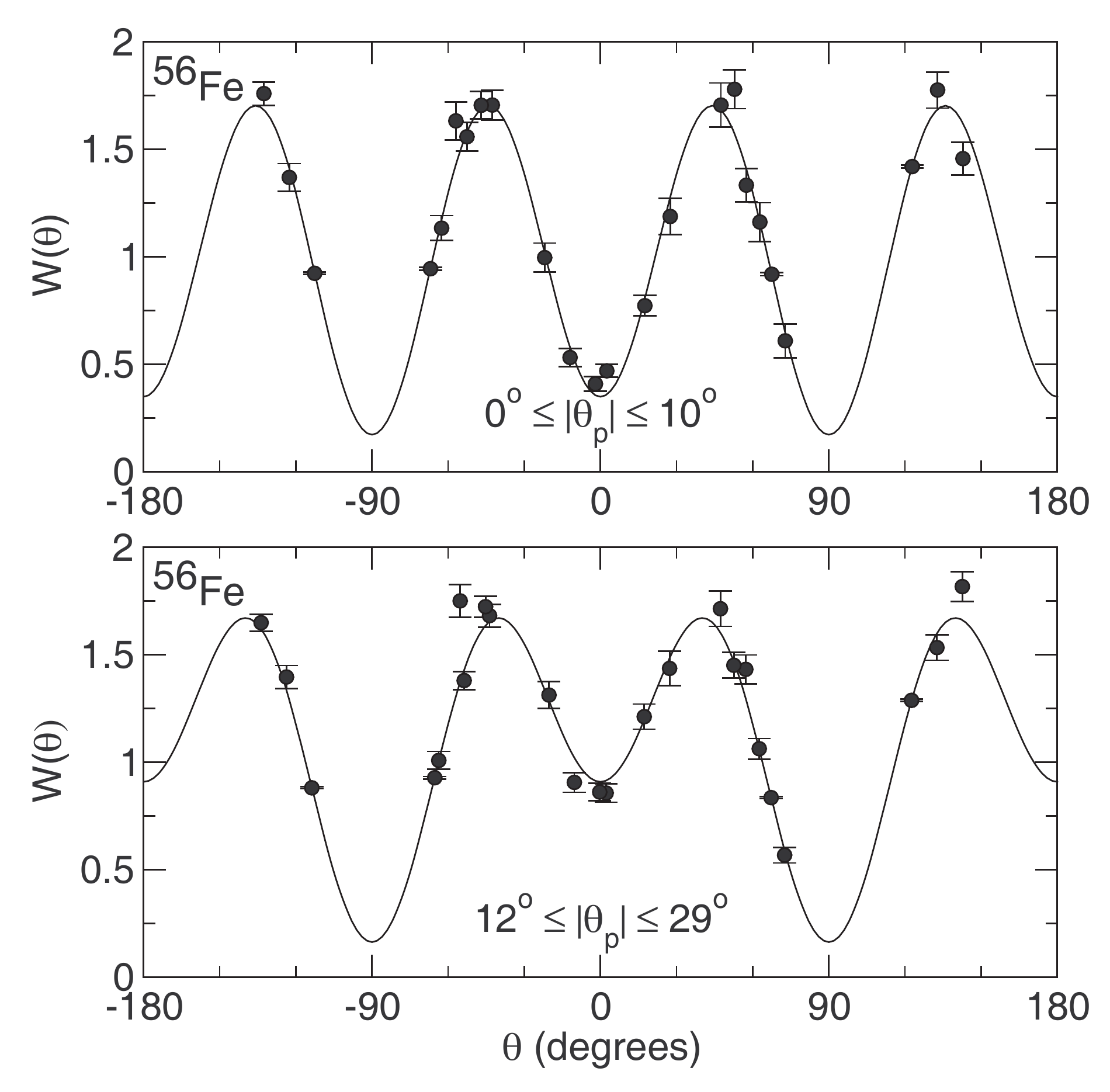}} \caption{Measured and calculated angular correlations for $^{56}$Fe in inverse kinematics from East {\em et al}. \cite{East.PhysRevC.79.024303}. The upper panel is for the centre
particle detector; the lower panel is for the top and bottom (outer) detectors. The range of particle detection angles in the vertical plane, above and below the beam axis, is indicated.}
\label{fig:Fe56AC}
\end{center}
\end{figure}

\begin{figure}
\begin{center}
\resizebox{8.5cm}{!}{\includegraphics{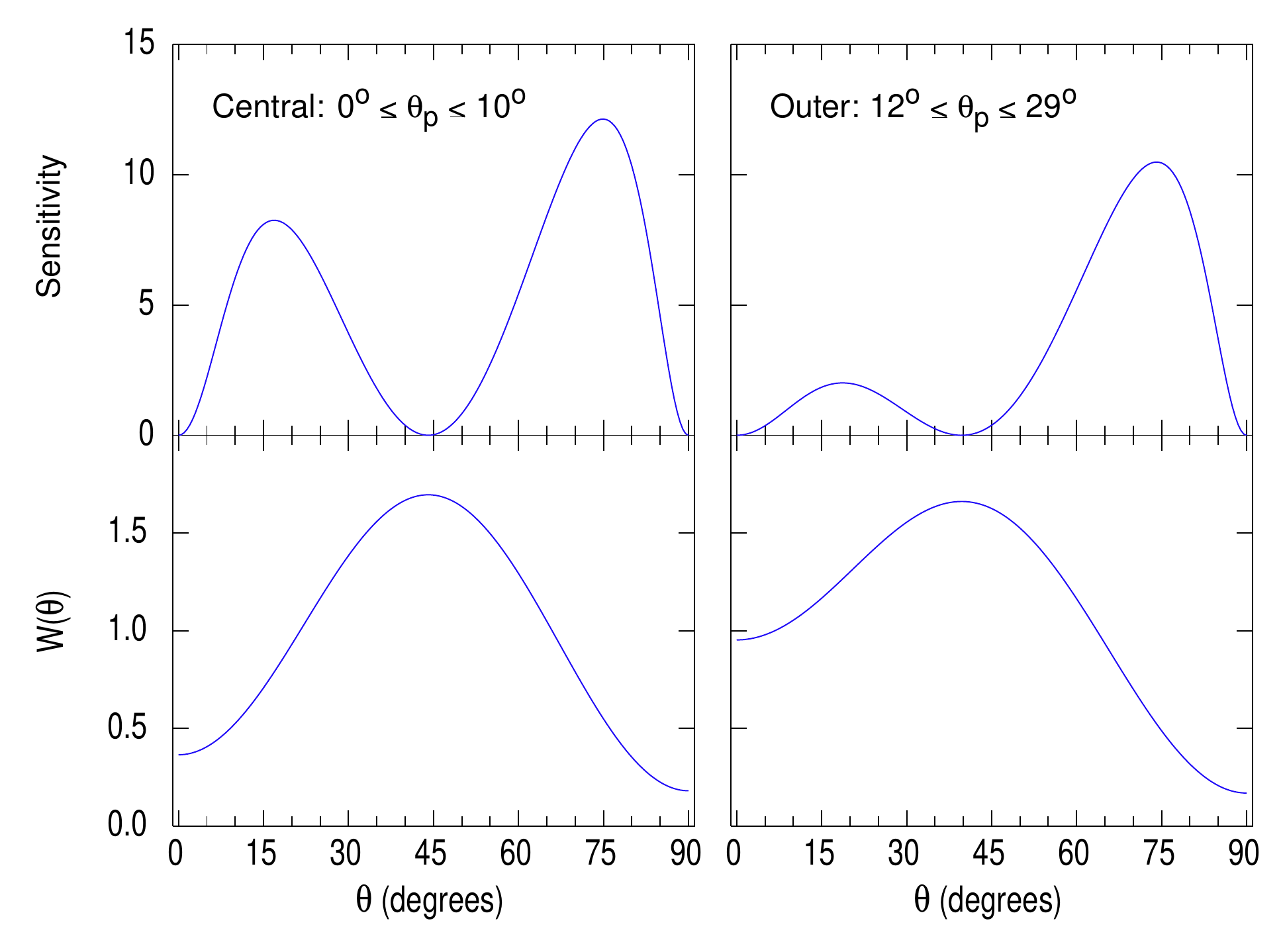}} \caption{Angular correlations and the corresponding sensitivity plots based on Eq.~(\ref{eq:S2N}) for the $2^+ \rightarrow 0^+$ transition of $^{56}$Fe shown in Fig.~\ref{fig:Fe56AC}, as a function of $\gamma$-ray detection angle. The left(right) panel is for the central(outer) particle detectors(s). The sensitivity here does not include the difference in excitation probability for the central and outer particle detectors.}
\label{fig:s2N-inverse}
\end{center}
\end{figure}

\subsection{Silicon photodiodes as particle detectors}

The silicon photodiode particle detectors are connected and biased via ORTEC 142B preamplifiers. Generally, when detecting heavy ions with energies of some tens of MeV, an energy signal will be obtained without applying bias. However a bias voltage is usually required to produce the fast timing output from the Ortec 142B preamplifier. Initially a bias voltage of $-20$~V is sufficient, but as the detectors become radiation damaged, the bias voltage must be increased. It can reach as high as $-150$~V before the detectors are replaced. As a practical indication of the damage rate, the PDB-C613-2 model photodiodes used to measure forward-scattered $^{26}$Mg in Ref.~\cite{MCCORMICK2018} were replaced after 48 hours of counting at $\sim 13$ kHz. For typical backscatter measurements, the detector count rates with typical beams of $\sim 1$~ pnA are usually about 1 kHz (often less) and the rate of radiation damage is very much lower. In this case the detectors can remain serviceable for several experimental runs of a week's duration.

The photodiodes are connected to the heat shield at $\sim 100$~K via electrical insulators, which are also poor thermal conductors. Presumably they equilibrate at the temperature near that of the first stage after a few hours. When so cooled the detectors continue to operate well with reduced leakage current. A cooled detector that has not been radiation damaged can resolve (not separate) the 5486- and 5443-keV $\alpha$ decays of $^{241}$Am. The spectrum at room temperature is practically the same as that of a (room temperature) surface barrier detector. For in-beam experiments with heavy ions, the particle energy spectra obtained with these photodiodes in transient-field $g$-factor measurements, whether conventional or inverse kinematics, have been of equal quality to those obtained previously with surface barrier detectors.

These photodiodes, which are not specifically designed to operate as heavy-ion detectors, show some variation in their behaviour. For example, the effective depletion depth, which seems to be typically of the order of 50 $\mu$m, appears to vary. This statement is based on a few observations of the energy at which the photodiodes begin to depart from behaving like a full-energy detector and begin to behave like a $\Delta E$ detector. A systematic study has not been made because in most cases the detected particles clearly stop within the depletion depth. In one case a series of proton beams with energies between 1.5 and 3.0 MeV were Rutherford scattered from a 1.5 mg/cm$^2$ thick Ta foil and observed simultaneously by a photodiode and a 100 $\mu$m surface barrier detector at back angles. The inference in that case was that the photodiode had an effective depletion depth of about 85 $\mu$m \cite{SRTee}.

\subsection{$g$-factor measurements in inverse kinematics}

The inset in Fig.~\ref{fig:trifid} illustrates the target structure and reaction geometry for a transient-field measurement in inverse kinematics. Whereas the ``conventional" kinematics measurements use a lighter beam on a heavier target isotope under study, and detect de-excitation $\gamma$~rays in coincidence with backscattered beam ions, in ``inverse" kinematics the beam ions are excited and de-excitation $\gamma$-rays are observed in coincidence with the lower-mass target ions that are knocked forward. In most cases three photodiodes arranged as shown in Fig.~\ref{fig:trifid} have been employed to detect the knocked-on target ions. See also Fig.~\ref{fig:closeup} which shows the detectors located in the target chamber.

The collisions in such inverse kinematics measurements are near head-on in the centre-of-mass frame; they are exactly head-on when the target ions strike the centre detector along the beam axis (see Fig.~\ref{fig:trifid}).

In the lab frame, the beam ion, which is considerably heavier than the target, always scatters close to the beam axis. However there can be quite some variation in the momentum transferred to the target ion, depending on the scattering angle. Target ions striking the centre detector are always higher in energy than those striking the outer detectors. In designing the experiment and the thickness of the target layers, it has to be considered whether the knocked-on target ions reaching the top and bottom extremes of the outer particle detectors will have sufficient energy to be detected above noise. For this reason it is best not to use an excessively thick copper backing layer to stop the (not scattered) beam. The problem then is that some beam ions might emerge from the target, and along with knock-on copper backing ions and delta electrons, strike the central particle detector causing high count rate and radiation damage. These are best blocked by covering the central detector (alone) by a mylar foil. Typically $\sim 1.7$~mg/cm$^2$ is sufficient.

Due to energy conservation, the angle-dependence of the energy of the knock-on target ions means that the velocity range of the scattered beam ions as they traverse the gadolinium layer of the target can be quite different for the central versus the outer detectors. The average beam-ion velocity is higher for knock-on target-ion events registered in the outer detectors than for those registered in the centre detector. Thus the transient field experienced by the ions differs somewhat for the centre and outer detectors. This difference is readily handled by analyzing the data separately for the centre and outer detectors. See Ref.~\cite{Chamoli.PhysRevC.83.054318} for an example and further details.

\begin{figure}
\begin{center}
\resizebox{7.0cm}{!}{\includegraphics{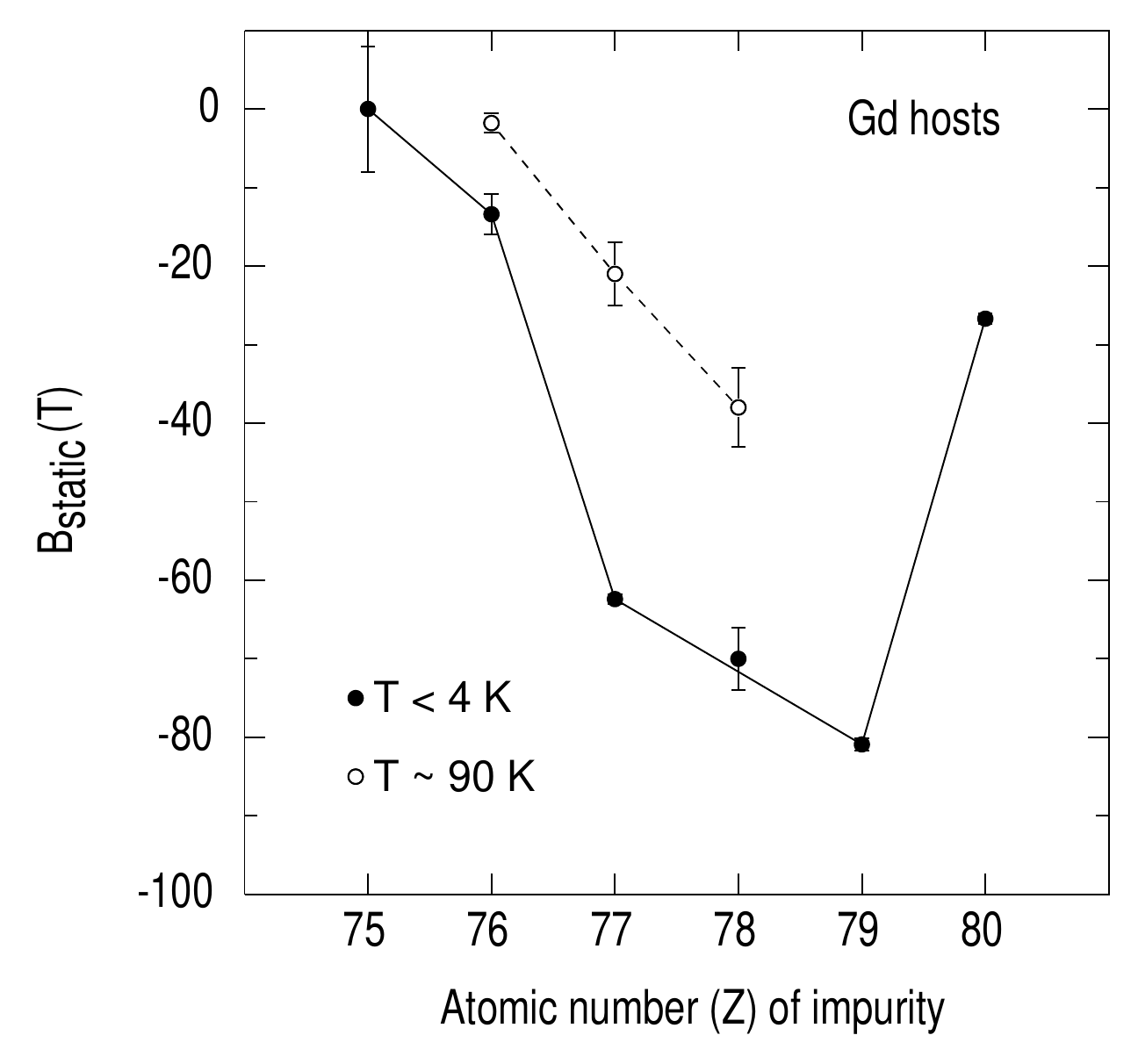}}
\caption{Hyperfine magnetic fields for impurities with $75 \geq Z \geq 80$ in Gd hosts. The data for temperatures below 4~K are from the compilation of Krane \cite{Krane1983}, apart from the Pt case which is from \cite{Carlsson1987,Mayer1988}. The data near 90 K are from Forker et al. (Os, Z = 76) \cite{Forker1978}, unpublished work from our laboratory (Ir, Z = 77) \cite{AES2000}, and Stuchbery and Anderssen (Pt, Z = 78) \cite{Stuch1995.PhysRevC.51.1017}. }
\label{fig:Bstatic}
\end{center}
\end{figure}

\begin{figure}
\begin{center}
\resizebox{7.0cm}{!}{\includegraphics{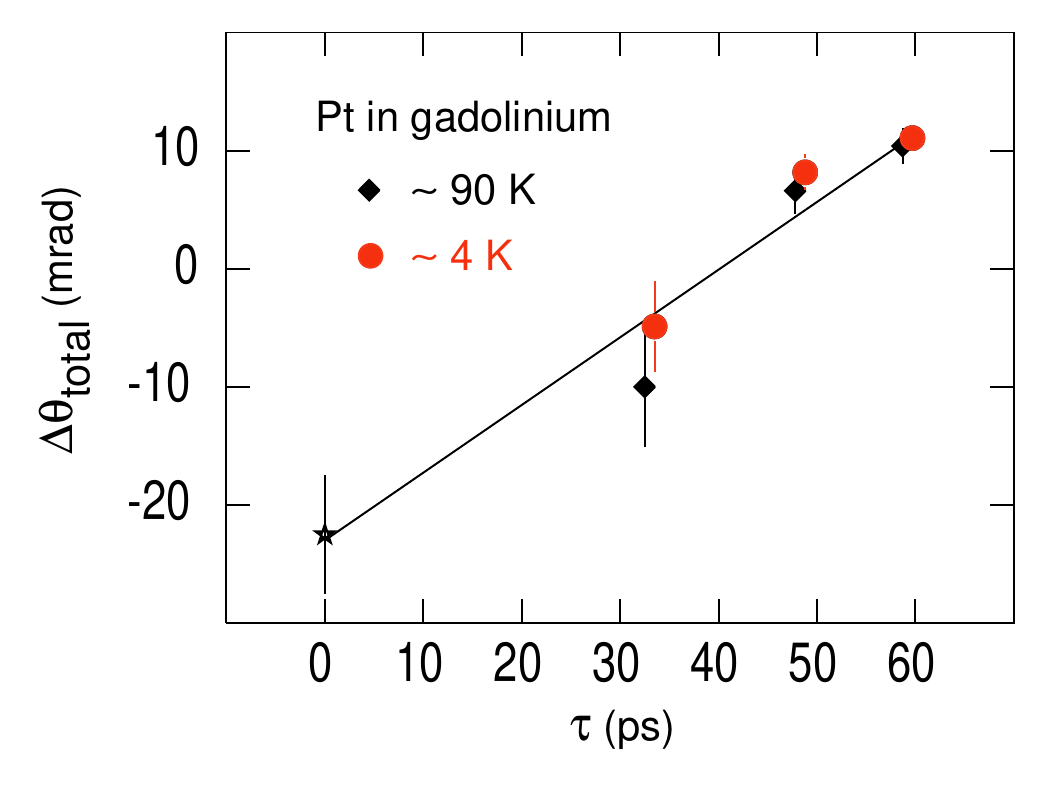}}
\caption{Precession angles for the first-excited 2$^+$ states of $^{194,196,198}$Pt measured  by the IMPAC method as a function of their mean lifetime $\tau$, at target temperatures of $\sim 90$~K (black diamonds) \cite{Stuch1995.PhysRevC.51.1017} and $\sim 4$~K (red circles). The line is a least squares fit to the data taken at $\sim 90$~K \cite{Stuch1995.PhysRevC.51.1017}. The $g$~factors of these 2$^+$states are the same within within uncertainties \cite{ANDERSSEN1995}, so the slope of the line determines the hyperfine field strength.
The transient-field precession angle, plotted at $\tau=0$, was calculated using the Rutgers parametrization \cite{Shu80.PhysRevC.21.1828} and allocated a 20\% uncertainty, as described in Ref.~\cite{Stuch1995.PhysRevC.51.1017}. These data indicate no significant temperature dependence below 90~K for the hyperfine field of Pt in gadolinium following implantation.
}
\label{fig:PtGdSFDtheta}
\end{center}
\end{figure}

\subsection{Optimal placement of detectors for inverse kinematics}

In section~\ref{sect:expt-design-opt}, a figure of merit was derived for perturbed angular correlation measurements, which can be used to optimize experimental conditions, including to determine the optimal angles at which to place the particle and $\gamma$-ray detectors. Examples were given for conventional kinematics in Fig.~\ref{fig:s2N}. Here we consider inverse kinematics, taking the case of $^{56}$Fe as reported by East {\em et al.} \cite{East.PhysRevC.79.024303}. A beam of $^{56}$Fe at an energy of 110 MeV was excited on a 0.6 mg/cm$^2$ C target. The particle-$\gamma$ angular correlations $W(\theta_{\gamma})$ measured in that work are shown in Fig.~\ref{fig:Fe56AC}. Figure~\ref{fig:s2N-inverse} shows the angular correlations and sensitivity as defined by Eq.~(\ref{eq:S2N}). In these measurements and calculations the $\gamma$-ray detector was placed to subtend an opening half angle of 18$^\circ$.

Comparing the central detector on the beam axis with the upper and lower detectors, it is evident that the shape of the angular correlations is similar, but there is less anisotropy at 0$^{\circ}$ and 180$^{\circ}$ for the outer detectors. However the slope at $\theta_{\gamma}=65^{\circ}$, i.e. $S(65^{\circ})$, as well as the sensitivity at $65^{\circ}$, is similar for the central and outer detectors.

It should be noted that the sensitivity evaluation in Fig.~\ref{fig:s2N-inverse} does not include an evaluation of the Coulomb-excitation cross section change between the central and the outer detectors. These relative cross sections vary from case to case. As a rough guide, the Coulomb-excitation cross sections for the central and outer detectors are often similar. This similarity comes about because the Coulomb-excitation cross section is the product of the Rutherford cross section and an excitation probability. The excitation probability is smaller for the outer detectors, but this is largely compensated for by an increase in the Rutherford cross section. In the example shown, the cross section for one of the outer detectors is about 90\% of that for the central detector.

\section{Commissioning experiments: temperature dependence of hyperfine fields}
\label{sect:commissioning}

The commissioning experiments were chosen to make use of and access the lower target temperatures and temperature control features of the Hyperfine Spectrometer that had not been available previously in our laboratory.

\subsection{Pt in gadolinium static hyperfine field}

The first experiment performed with the Hyperfine Spectrometer was an IMPAC measurement to study the static hyperfine field at Pt isotopes implanted into a polarized gadolinium foil held at a nominal temperature of $\sim 4$~K. The states of interest in a natural Pt target were Coulomb exited by a 36-MeV $^{16}$O beam and recoil implanted into the gadolinium backing layer.  A similar measurement had previously been performed at $\sim 90$~K using LiN$_2$ for cooling the target~\cite{Stuch1995.PhysRevC.51.1017}.  As indicated in Fig.~\ref{fig:Bstatic}, the static-field strength of $-38\pm5$~T observed was smaller by about a factor of two than expected based on systematics of measurements performed at temperatures near 4~K. There are three possible reasons for this difference: (i) The static hyperfine field may vary rapidly with temperature, departing from the temperature dependence of the host magnetization. Such behaviour has been observed for Os in gadolinium \cite{Forker1978}. (ii) The ion implantation process could be disruptive and result in many Pt nuclei on low-field sites, thus reducing the average hyperfine field. (iii) The applied polarizing field of $\sim 0.1$~T might not be sufficient to observe the full hyperfine field strength \cite{Stuch98NMM}.

As the first of these possibilities was proposed in Ref.~\cite{Stuch1995.PhysRevC.51.1017}, it was investigated first by repeating the IMPAC measurement at the lowest available temperature. Results are shown in Fig.~\ref{fig:PtGdSFDtheta}. There is no detectable difference between the hyperfine fields observed in IMPAC measurements performed at target temperatures of $\sim 90$~K and $\sim 4$~K (this is the nominal temperature of the target - that in the beam spot could be somewhat higher but it does not affect this conclusion or the following discussion).

\begin{table}[t]
  \centering
  \begin{threeparttable}
\caption{Transient-field $g$-factor measurements with the Australian National University Hyperfine Spectrometer.}
\label{tab:TFmeas}
\begin{tabular}{lll}
\toprule
Nuclide & Host & Reference\\
\midrule

\multicolumn{2}{l}{Conventional kinematics \tnote{1}}\\

$^{100,102,104}$Ru  & \textit{Fe}  & \cite{Chamoli.PhysRevC.83.054318} \\

$^{110,112,114,116}$Cd  & \textit{Fe}  & \cite{Chamoli.PhysRevC.83.054318}  \\

$^{111,113}$Cd  & \textit{Fe}  & \cite{Stuchbery.PhysRevC.93.031302} \\

$^{111}$Cd  & \textit{Fe}  & \cite{Coombes2019} \\ 

$^{116,118,120}$Sn & \textit{Fe}  & \cite{EAST2008} \\ 

$^{125}$Te  & \textit{Fe}  & \cite{Chamoli.PhysRevC.80.054301} \\ 

$^{124,126,128,130}$Te  & \textit{Fe}  & \cite{Coombes2019a} \\ 
\\

\multicolumn{2}{l}{Inverse kinematics \tnote{2}}\\

$^{56,57}$Fe  & \textit{Gd}   & \cite{East.PhysRevC.79.024303} \\ 

$^{54,56,58}$Fe  & \textit{Gd}   & \cite{East.PhysRevC.79.024304} \\ 

$^{70,72,74,76}$Ge & \textit{Fe}   & \cite{McCormick2019} \\ 

$^{74,76,78,80,82}$Se & \textit{Fe} and \textit{Gd} & \cite{McCormick2019} \\ 

$^{77}$Se & \textit{Gd} &  \cite{AES2012} \\ 

$^{96,98}$Mo  & \textit{Gd} & \cite{Chamoli.PhysRevC.83.054318} \\

$^{96,98,100,102,104}$Ru & \textit{Gd} & \cite{Chamoli.PhysRevC.83.054318} \\

$^{102,104,106,108,110}$Pd  & \textit{Gd} & \cite{Chamoli.PhysRevC.83.054318} \\

$^{106,108,112,114}$Cd  & \textit{Gd} & \cite{Chamoli.PhysRevC.83.054318} \\
\\

\multicolumn{2}{l}{Projectile excitation \tnote{3}}\\

$^{24,26}$Mg & \textit{Gd} &  \cite{MCCORMICK2018} \\ 

      \bottomrule
    \end{tabular}
    \begin{tablenotes}
      \item[1]{Target excitation by lower-mass beam. Beam ions detected at back angles.}
      \item[2]{Projectile excitation on lower-mass target. Target ions detected at forward angles.}
      \item[3]{Projectile excitation on higher-mass target. Beam ions detected at forward angles.}
    \end{tablenotes}
\end{threeparttable}
\end{table}

After publication of Ref.~\cite{Stuch1995.PhysRevC.51.1017} we became aware of a conference paper \cite{Carlsson1987} and unpublished report \cite{Mayer1988} on $^{192}$Pt in gadolinium by the Integral Perturbed Angular Correlation (IPAC) radioactivity method. The hyperfine field at 30~K is $B_{\rm static}=-70(4)$~T, which agrees with the low-temperature systematics in Fig.~\ref{fig:Bstatic}.
The fact that the IPAC data were taken with external fields up to 3.3~T, whereas the IMPAC data have a much smaller polarizing field of $\sim 0.1$~T, may play a part in explaining the difference in hyperfine field strengths; however the IPAC measurements use spherical samples whereas the IMPAC use thin foils, which are easier to magnetize.

Since no temperature dependence was observed in the IMPAC data, the difference between the IMPAC data and other methods is likely to relate to the implantation process rather than to the temperature at which the measurement is performed. Many implanted nuclei appear to be on low-field sites. This hypothesis could be tested by time dependent perturbed angular correlation measurements, which can identify whether there is a unique field or a distribution of hyperfine fields at the implantation site(s) (see section~\ref{sect:TDPAD} below).

\subsection{Temperature dependence of transient fields}

A second group of commissioning experiments used the ability to vary and control the target temperature between $\sim 4$~K and room temperature. The temperature dependence of the transient field acting on $^{196}$Pt ions traversing gadolinium and terbium hosts was measured. Some of these results have been shown above in Fig.~\ref{fig:magGd} and Fig.~\ref{fig:magnetizn}.

\section{Transient-field $g$-factor measurements}
\label{sect:TF}

Since the commissioning phase, the Hyperfine Spectrometer has been employed primarily to perform transient-field $g$-factor measurements following Coulomb excitation. Table~\ref{tab:TFmeas} summarizes those performed to date, including some that have yet to be published. The experiments include measurements in conventional kinematics (target excitation), inverse kinematics (beam excitation on a lower mass target) and the case of $^{24,26}$Mg where the projectiles were excited on a gadolinium foil that served as both the target and the ferromagnetic host.

\begin{figure}
\begin{center}
\resizebox{7.0cm}{!}{\includegraphics{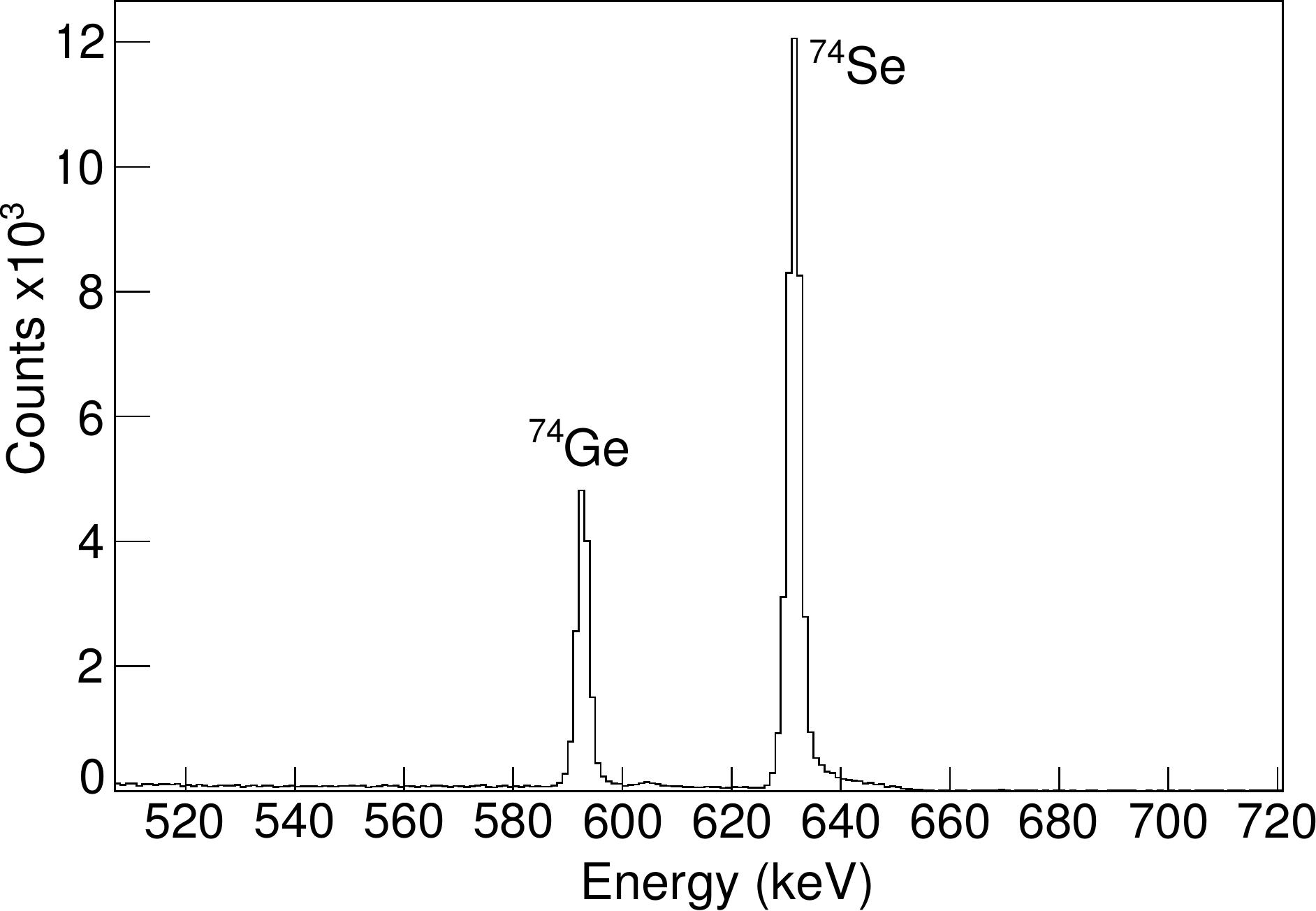}}
\caption{Spectrum of $\gamma$ rays observed at 65$^{\circ}$ to the beam in coincidence with forward recoiling $^{12}$C target ions during a transient-field $g$-factor measurement in inverse kinematics with a cocktail beam of $^{74}$Ge and $^{74}$Se. See McCormick {\em et al.} \cite{McCormick2019} for additional details.}
\label{fig:SeGe74}
\end{center}
\end{figure}

For the $g$-factor measurements, the cryocooler has been run without controlling the temperature, so that the target typically stays near 5~K, depending on the heat load. Such cooling power is not needed for either iron or gadolinium hosts (see. Fig.~\ref{fig:magGd}), however it has proved to be of great advantage for measurements on low melting point (MP) targets like Cd (MP 321$^{\circ}$C) and Sn (MP 232$^{\circ}$ C). High beam currents (several particle nanoamperes) could be maintained on these targets for several days without evidence of deterioration.

Many of the measurements performed have used inverse kinematics. The study of the Ge and Se isotopes included the first use of a cocktail beam to measure the transient-field precessions of excited states in $^{74}$Ge and $^{74}$Se simultaneously - thus avoiding the majority of the possible systematic errors that could affect sequential measurements. An example of a $\gamma$-ray spectrum obtained with the cocktail beam is shown in Fig.~\ref{fig:SeGe74}.

\begin{figure*}
\begin{center}
\resizebox{12.0cm}{!}{\includegraphics{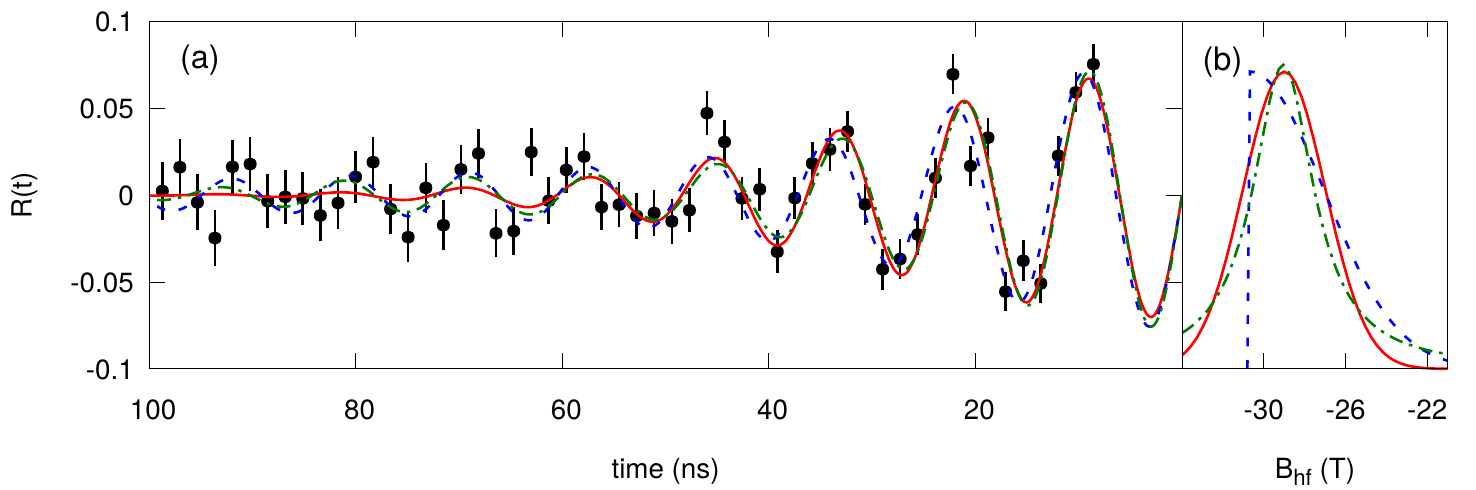}}
\caption{TDPAD ratio function for the $11/2^-$ isomeric state ($E_x=846$~keV, $\tau=107$~ns) of $^{107}$Cd implanted into gadolinium compared with simulations for different field distributions: red solid line, Gaussian; blue dashed line, half-Gaussian; green dashed-dotted
line, Lorentzian. (a) Fits of the ratio functions. (b) Field distributions. See Ref.~\cite{Gray.PhysRevC.96.054332} for further details.
}
\label{fig:CdGdTDPAD}
\end{center}
\end{figure*}

\section{Time dependent perturbed angular distributions}
\label{sect:TDPAD}

The pulsed beams available from the Heavy Ion Accelerator Facility combined with the capabilities of the Hyperfine Spectrometer allow studies of the temperature dependence of hyperfine fields following ion implantation by time dependent perturbed angular distribution (TDPAD) methods \cite{Raghavan1985}. In addition, the development of lanthanum bromide (LaBr$_3$) scintillator detectors provides an opportunity to perform these experiments under new experimental conditions through their combination of good energy and time resolution. As a first application of LaBr$_3$ detectors to in-beam TDPAD techniques, the hyperfine field of Cd implanted into gadolinium was investigated \cite{Gray.PhysRevC.96.054332} and related to an earlier measurement of the $g$-factor of the $I^{\pi} = 10^+$ state in $^{110}$Cd by an integral IMPAD method \cite{REGAN1995}. Figure \ref{fig:CdGdTDPAD} shows an example of a TDPAD ratio function for $^{107}$Cd implanted into gadolinium at a nominal temperature of 6~K following the $^{98}$Mo($^{12}$C, $3n$)$^{107}$Cd reaction at a beam energy of 48 MeV. The attenuation of the amplitude of the oscillations is attributed to there being a range of implantation sites with differing hyperfine field strengths.

Some additional exploratory studies of this type have been performed with a view to both understanding the hyperfine fields following ion implantation, and to performing TDPAD $g$-factor measurements in-beam on states that have not previously been accessible due to high precession frequencies and/or short lifetimes \cite{gray2019}.


\section{Conclusions and future applications}

The design and operation of the ANU Hyperfine Spectrometer, an apparatus for in-beam hyperfine interactions studies and $g$-factor measurements, has been described. To date, the Hyperfine Spectrometer has been used mainly for transient-field $g$-factor measurements; however it was designed for a wide range of studies at the interface of nuclear and condensed matter physics, which may be explored in future work.

As a first example of proposed future research, the after-effects of ion implantation on a time-scale of picoseconds can be studied via observations of pre-equilibrium effects in hyperfine fields \cite{Stuchbery.PhysRevLett.82.3637,Anderssen1995HFI,Stuchbery1996,Chien.PhysRevB.19.1363,Alfter1997}. Open questions concern the extent to which the properties of the host affect the equilibration process and whether there is any significant component in the equilibration time that should be associated with the interactions that generate the hyperfine field \cite{Alfter1997} rather than with the implantation mechanism \cite{Stuchbery.PhysRevLett.82.3637,Anderssen1995HFI,Stuchbery1996,Chien.PhysRevB.19.1363}. These questions can be addressed by studying the equilibration time for several hosts as a function of temperature.

For a second example of future work, the availability of LaBr$_3$ detectors has opened the possibility to measure the magnetic moments of relatively short-lived isomers ($\tau > \sim 10$~ns) by time dependent perturbed angular distribution (TDPAD) methods following heavy ion reactions and recoil implantation into ferromagnetic hosts. Some preliminary studies of gadolinium hosts have been completed \cite{Gray.PhysRevC.96.054332,gray2019}. In addition to gadolinium, the rare earth metals terbium, dysprosium, holmium, and erbium all become ferromagnetic at temperatures between 20~K and 220~K. There is therefore the possibility to exploit the hyperfine fields in these ferromagnetic materials for nuclear moment measurements. On one hand a series of conceptually simple experiments may be possible, where the one foil serves as both the target material and the ferromagnetic host. On the other hand, the hyperfine fields in these hosts are not well known, and their crystal structures are such that electric field gradients will be present along with the magnetic hyperfine fields. Exploratory studies will be necessary. Nevertheless, Alfter {\em et al.} \cite{Alfter1995} have reported measurements in this spirit, albeit with neutron activation and integral perturbed angular correlation measurements on short-lived states. Time-dependent measurements can make it possible to disentangle the contributions to the overall hyperfine interaction that are hidden in integral measurements.

\section*{Acknowledgments}

The Hyperfine Spectrometer was funded by an Australian Research Council Research (ARC), Infrastructure Equipment and Facilities Grant (2001) together with an Australian National University Major Equipment Grant. Research using the Hyperfine Spectrometer has been supported by ARC Discovery grants numbers DP0773273 and DP170101673. The authors thank the academic and technical staff of the Department of Nuclear Physics (Australian National University) and the Heavy Ion Accelerator Facility for their support. The contributions of many colleagues and students to the development and execution of the research program are gratefully acknowledged. The contributions of Dr Anna N. Wilson, Dr Paul M. Davidson, and Michael C. East at the commissioning stage are particularly acknowledged. Dr No\'emie Koller (Rutgers University) is thanked for performing the magnetization measurements on the gadolinium and terbium foils and for many inspiring discussions. Support for the ANU Heavy Ion Accelerator Facility operations through the Australian National Collaborative Research Infrastructure Strategy (NCRIS) program is acknowledged.

\appendix
\section{Evaluating angular correlations for rectangular particle detectors}
\label{sect:rectangular_tensors}

Experience over several decades has shown that for transitions of known multipolarity (usually pure $E2$ transitions or mixed $E2/M1$ transitions with known mixing ratio), the particle-$\gamma$ angular correlations after Coulomb excitation can be calculated accurately. This experience was initially accumulated using annular surface barrier detectors in backscatter (or conventional kinematics) geometry  \cite{BOLOTIN1983,Stuchbery1985,STUCHBERY1985-Os,BYRNE1987,STUCHBERY1988,LAMPARD1989,Doran.PhysRevC.40.2035,Stuchbery1991,STUCHBERY1991a,Stuchbery1992,Stuchbery1994,LAMPARD1994,ANDERSSEN1995,STUCHBERY1998,ROBINSON1999,STUCHBERY2000,BEZAKOVA2000,Mantica.PhysRevC.63.034312}. This appendix summarizes some technical aspects of evaluating statistical tensors and angular correlations for rectangular particle detectors in Coulomb excitation experiments, based on the de Boer-Winther code~\cite{wdb}. The difference between the evaluation for annular counters and rectangular photodiode detectors is in the evaluation of the statistical tensor that specifies the spin alignment of the initial state.

\subsection{Statistical tensors}

Figure~\ref{fig:particle-angles} shows the co-ordinate system used to evaluate the statistical tensors and the angular correlations. The beam direction defines the $z$-axis, the magnetic field direction along the $y$-axis, and the $\gamma$-ray detectors are in the $xz$-plane. Thus the $\gamma$-ray detectors are in the $\phi = 0$ plane. The angular correlation is given by Eq.~(\ref{eq:pac}). To evaluate the angular correlations relevant to the experiment requires evaluation of the statistical tensors that define the alignment of the nuclear state of interest. For this purpose we begin with the statistical tensors calculated by the de Boer-Winther code, which are evaluated in their co-ordinate frame 3 with the beam along the $z$-axis and with the {\em reaction} in the $xz$ or $\phi=0$ plane (i.e. $\phi=0$ is {\em not} the $\gamma$-ray detection plane in this calculation).

The statistical tensors $B_{kq}$ in the notation of Eq.~(\ref{eq:pac}) are related to those of de Boer and Winther, $\alpha_{kq}^{(3)}$, by
\begin{equation}
B_{kq} = \sqrt{2k+1} \; \rho_{kq} = \sqrt{2k+1} \; \frac{\alpha_{kq}^{(3)}}{\alpha_{00}^{(3)}}.
\end{equation}
In the co-ordinate frame of the experiments,  Fig.~\ref{fig:particle-angles}, tensors for scattering at angle $\phi_p$ are given by a rotation through $\phi_p$ about the beam axis:
\begin{eqnarray}
B_{kq}(\theta_p, \phi_p) &=& \sum_{q^\prime} B_{k
q^\prime}(\theta_p, 0) D^k_{q^\prime q}(\phi_p,0,0) \\
 &=& B_{k q}(\theta_p, 0)
 {\rm e}^{i q\phi_p}.
\end{eqnarray}

This statistical tensor must be averaged over the scattered particle angles relevant for the particle detector(s) and over the energy loss of the beam in the target:
\begin{equation}
\label{eq:avtens}
\langle B_{kq} \rangle = \frac{\int_{E} \int_{\theta_p} \int_{\phi_p} B_{k q}(\theta_p, 0) {\rm e}^{i q\phi_p} \frac{{\rm d}^2 \sigma}{{\rm d} \Omega_p {\rm d} E} {\rm d} \Omega_p {\rm d}E }
 {\int_{E} \int_{\theta_p} \int_{\phi_p}  \frac{{\rm d}^2 \sigma}{{\rm d} \Omega_p {\rm d} E} {\rm d} \Omega_p {\rm d}E},
\end{equation}
where $\frac{{\rm d}^2 \sigma}{{\rm d} \Omega_p {\rm d} E}$ is the Coulomb excitation cross section as a function of scattering angle and beam energy.

For the case of two identical detectors placed equidistant from the beam axis, above and below it as in Fig.~\ref{fig:particle-angles}, or for a detector centred on the beam axis as in Fig.~\ref{fig:trifid}, the integration over the particle detector acceptance can be limited to the positive quadrant specified by $0 \leq \phi_p \leq \pi/2$ by replacing the factor ${\rm e}^{i q\phi_p} $ in Eq.~(\ref{eq:avtens}) by $( {\rm e}^{iq \phi} + {\rm e}^{-iq \phi} + {\rm e}^{iq (\phi+\pi)} + {\rm e}^{-iq (\phi+\pi)} )/4$, which is $\cos qx$ if $q$ is even and is zero if $q$ is odd.


\subsection{Solid angle in Cartesian co-ordinates}

The solid angle is usually given in spherical polar coordinates as
\begin{equation}
{\rm d} \Omega = \sin \theta {\rm d} \theta {\rm d} \phi .
\end{equation}
(The formulae here are general so the subscript $p$ used above for the particle-detector solid angle is now omitted.) To evaluate the cross sections and statistical tensors for rectangular particle detectors it is advantageous to use Cartesian co-ordinates. In this case the solid angle is
\begin{equation}
{\rm d} \Omega = \frac{\cos \vartheta}{r^2} {\rm d} x {\rm d} y ,
\end{equation}
where $\vartheta$ is the angle between the direction of the vector
$\bm{r} = x{\bm i} + y{\bm j} + z {\bm k}$ from the position of the beam on the target (i.e. the origin) to the point $(x,y,z)$ on the particle detector and the vector which specifies the direction
normal to the element of surface at the point $(x,y,z)$. Thus when this normal is parallel to the beam (or $z$) axis,
\begin{equation}
{\rm d} \Omega = {z \: {\rm d} x {\rm d} y} / {|r^3|}.
\end{equation}


\subsection{Statistical tensors in inverse kinematics}

The statistical tensors in the case of inverse kinematics (projectile excitation on a lower-mass target) have been evaluated by transforming to the centre of mass frame and evaluating the tensors and cross sections for projectile excitation with the de Boer-Winther code (in the centre-of-mass frame), before transforming back to the lab frame.


\subsection{Evaluating the angular correlation}

Once the statistical tensors are known, the angular correlation can be evaluated from Eq.~(\ref{eq:pac}). The solid angle attenuation factors, $Q_k$, can be evaluated with sufficient precision by using the method of Krane \cite{KRANE1972-205,KRANE1973-401}. The $F_k$ coefficients have been tabulated by Yamazaki \cite{YAMAZAKI1967} or may be computed readily. Likewise, routines are available to evaluate the Wigner rotation matrices; they can also be replaced by spherical harmonics in Eq.~(\ref{eq:pac}) by using
\begin{equation}
D^k_{q0}(\alpha,\beta,\gamma) = \sqrt{\frac{4 \pi}{2k+1}} Y_{kq}(\beta,\alpha).
\end{equation}

Figure~\ref{fig:Fe56AC} shows one example of a comparison of the calculated angular correlation with experiment. In this case there are no free parameters in the fit to experimental data, apart from an overall normalization factor.

\subsection{Legendre polynomial expansion}

The theoretical angular correlation for a rectangular particle detector is a sum over Wigner rotation matrices, $D^k_{q0}$; see Eq.~(\ref{eq:pac}). However, as noted above, these have a zero index and can be reduced to spherical harmonics. In addition, if the $\gamma$-ray detectors are in the $\phi=0$ plane, the spherical harmonics reduce to associated Legendre polynomials. It is sometimes more convenient, however, to fit the experimental data using ordinary Legendre polynomials. For cases like those illustrated in Fig.~\ref{fig:particle-angles} with the $\gamma$-ray detection in the $xz$ or horizontal plane, and the particle detectors symmetrically placed vertically above and below this plane, as well as being symmetrical left-right of the vertical (or $y$) axis, Eq.~(\ref{eq:pac}) reduces to
\begin{equation}
W(\theta_\gamma) = \sum_{k q} B_{k q}(\theta_p) Q_k F_k D^{k *}_{q 0}(0, \theta_\gamma,0).
\label{eq:pac1}
\end{equation}
This expression can be rewritten in form of the familiar expansion in terms of Legendre polynomials
\begin{equation}\label{eq:Pk}
W(\theta_\gamma) =  A_0 + A_2 P_2(\cos \theta_\gamma) + A_4 P_4(\cos \theta_\gamma),
\end{equation}
by making use of the relations
\begin{eqnarray}
D^2_{00}(0,\theta,0) &=& P_2(\cos\theta) \\
D^2_{20}(0,\theta,0) &=& [1-P_2(\cos\theta)]/\sqrt{6} \\
D^4_{00}(0,\theta,0) &=& P_4(\cos\theta)\\
D^4_{20}(0,\theta,0) &=& \sqrt{\frac{2}{5}}[-P_4(\cos\theta)  \nonumber \\
                     &+& \frac{5}{6} P_2(\cos\theta) + \frac{1}{6}] \\
D^4_{40}(0,\theta,0) &=&  \sqrt{\frac{1}{70}}  [ P_4(\cos\theta)   \nonumber \\
                      &-& \frac{10}{3} P_2(\cos\theta) + \frac{7}{3}  ].
\end{eqnarray}
The result is
\begin{eqnarray}
A_{0} &=&1+\sqrt{\frac{2}{3}} B_{22} f_2   \nonumber \\
      &+ &  [\sqrt{\frac{2}{45}} B_{42} + \sqrt{\frac{14}{45}}  B_{44}] f_4 \\
A_{2} &=& [B_{20} -\sqrt{\frac{2}{3}} B_{22}] f_2 \nonumber \\
      &+& [\sqrt{\frac{10}{9}} B_{42} -\sqrt{\frac{40}{63}} B_{44} ] f_4 \\
A_{4} &=& [ B_{40} -\sqrt{\frac{8}{5}} B_{42} +\sqrt{\frac{2}{35}} B_{44} ]f_4,
\end{eqnarray}
where, for brevity, $f_k = F_k Q_k$. Of course the statistical tensors $B_{kq}$ must be evaluated in the same co-ordinate frame as shown in Fig.~\ref{fig:particle-angles}. Note that $A_0 \ne 1$ and that it depends on quantities normally associated with the $k=2$ and $k=4$ terms; likewise $A_2$ includes terms normally associated with $k=4$. Empirical fits to angular correlation data can still be performed with $A_0$, $A_2$, and $A_4$ as free parameters in Eq.~(\ref{eq:Pk}), but care is required in the interpretation of these parameters.

\section*{References}

\bibliography{AESTF_NIM}



\end{document}